\documentclass[longauth]{aa}

\usepackage{graphicx}
\usepackage{natbib}
\usepackage{flexisym}

\usepackage{letltxmacro}
\usepackage{longtable}
\usepackage{amsmath}
 
\usepackage{caption}
\usepackage{xcolor}
\LetLtxMacro\FlexiSymTextPrime\textprime
\DeclareTextCommandDefault{\textprime}{\FlexiSymTextPrime}  
\usepackage{subcaption}  
\usepackage{float}       
\usepackage{needspace}
\usepackage{graphicx}
\usepackage{caption}
\usepackage{subcaption}

\usepackage{longtable}
\usepackage{booktabs}
\usepackage{pgfplotstable}
\usepackage{siunitx}

\usepackage{hyperref}
\hypersetup{colorlinks=True, linkcolor=blue, citecolor=blue}

\usepackage{txfonts,textcomp}

\begin{document}

   \title{Resolving the extended interstellar medium of infrared galaxies II. Fine-structure line emission}

    \author{C. M. Lofaro\inst{1,2}, T. Díaz-Santos\inst{2,3}, L. Armus\inst{4}, V. Charmandaris\inst{1,2,3}, T. Lai\inst{4}, M. Pereira-Santaella\inst{5}, J. McKinney\inst{6}, R. Fernández-Aranda\inst{1,2,7}, G. Donnelly\inst{8}, M. Bianchin\inst{9,10}, Y. Song\inst{11,12}, I. Shivaei\inst{7}, L. A. Boogaard\inst{13}, G. Rodighiero\inst{14,15}, S. Linden\inst{16}, G. C. Privon\inst{17,18,19}, C. Ricci\inst{20,21}, L. Lenkić\inst{4}, H. Inami\inst{23}, S. Stierwalt\inst{24}, V. U\inst{4,25}, D. Kakkad\inst{26}, P. Van der Werf\inst{13}, M. Sánchez-García\inst{2}, A. Evans\inst{17,19}, and D. B. Sanders\inst{27}}

    \institute{
    Department of Physics, University of Crete, 70013 Heraklion, Greece\\
    \email{clofaro@ia.forth.gr}
    \and
    Institute of Astrophysics, Foundation for Research and Technology--Hellas (FORTH), Heraklion, GR-70013, Greece
    \and
    School of Sciences, European University Cyprus, Diogenes street, Engomi, 1516 Nicosia, Cyprus
    \and
    IPAC, California Institute of Technology, 1200 E. California Boulevard, Pasadena, CA 91125, USA
    \and
    Instituto de Física Fundamental, CSIC, Calle Serrano 123, 28006 Madrid, Spain
    \and
    Department of Astronomy, The University of Texas at Austin, Austin, TX, USA
    \and
    Centro de Astrobiología (CAB), CSIC-INTA, Carretera de Ajalvir km 4, Torrejón de Ardoz, E-28850, Madrid, Spain
    \and
    Ritter Astrophysical Research Center and Department of Physics and Astronomy, University of Toledo, Toledo, OH 43606, USA
    \and
    Instituto de Astrofisica de Canarias, Calle Via Lactea, s/n, E-38205, La Laguna, Tenerife, Spain
    \and
    Departamento de Astrofisica, Universidad de La Laguna, E-38206, La Laguna, Tenerife, Spain
    \and
    European Southern Observatory, Alonso de Córdova, 3107, Vitacura, Santiago, 763-0355, Chile
    \and
    Max-Planck-Institut f\"{u}r Radioastronomie, Auf dem H\"{u}gel 69, 53121, Bonn, Germany
    \and
    Leiden Observatory, Leiden University, PO Box 9513, 2300 RA Leiden, The Netherlands
    \and
    Dipartimento di Fisica e Astronomia, Università di Padova, Vicolo dell’Osservatorio 3, 35122 Padova, Italy
    \and
    INAF-Osservatorio Astronomico di Padova, Vicolo dell’Osservatorio 5, 35122 Padova, Italy
    \and
    Steward Observatory, University of Arizona, 933 North Cherry Avenue, Tucson, AZ 85721, USA
    \and
    National Radio Astronomy Observatory, 520 Edgemont Road, Charlottesville, VA 22903, USA
    \and
    Department of Astronomy, University of Florida, P.O. Box 112055, Gainesville, FL 32611, USA
    \and
    Department of Astronomy, University of Virginia, 530 McCormick Road, Charlottesville, VA 22903, USA
    \and
    Department of Astronomy, University of Geneva, ch. d'Ecogia 16, 1290, Versoix, Switzerland
    \and
    Instituto de Estudios Astrofisicos, Facultad de Ingenieria y Ciencias, Universidad Diego Portales, Av. Ejercito Libertador 441, Santiago, Chile
    \and
    Kavli Institute for Astronomy and Astrophysics, Peking University, Beijing 100871, China
    \and
    Hiroshima Astrophysical Science Center, Hiroshima University, 1-3-1 Kagamiyama, Higashi-Hiroshima, Hiroshima 739-8526, Japan
    \and
    Occidental College, Physics Department, 1600 Campus Road, Los Angeles, CA 90042, USA
    \and
    Department of Physics and Astronomy, 4129 Frederick Reines Hall, University of California, Irvine, CA 92697, USA
    \and
    Centre for Astrophysics Research, Department of Physics, Astronomy and Mathematics, University of Hertfordshire, Hatfield, AL10 9AB, UK
    \and
    Institute for Astronomy, University of Hawaii, 2680 Woodlawn Drive, Honolulu, HI 96822, USA
    }

  \date{Received 7 February 2026 / Accepted 10 September 2026}

   \authorrunning{Lofaro C. M. et al.}

\abstract
   {Understanding how active galactic nuclei (AGNs) and starburst activity shape the interstellar medium (ISM) in luminous infrared galaxies (LIRGs) requires spatially resolved spectral diagnostics that probe the multiphase medium. In paper I, we constrained the properties of polycyclic aromatic hydrocarbon emission. In this work, we focus on the characterization of the ISM ionized and neutral-gas components.}
   {We quantify radial variations in dust grain and gas properties via infrared fine-structure line ratios. Our goal is to constrain the ionization parameter U, the hydrogen density $\rm n_{\mathrm{H}}$, the stellar age, and the metallicity in a representative sample of 67 nearby (U)LIRGs from the Great Observatories All-sky LIRG Survey (GOALS) sample.}
   {We combined mid-infrared (MIR) spectral mapping from the Infrared Spectrograph (IRS) on board the Spitzer Space Telescope with far-infrared (FIR) spectroscopy from the Photodetector Array Camera and Spectrometer (PACS) on board the Herschel Space Observatory to perform a spatially resolved analysis of the galaxies on kiloparsec scales. Fine-structure emission lines were measured using different aperture grids extending beyond the nuclear regions up to $\sim$14 kpc away from the galaxies' nuclei. The observed MIR and FIR line ratios were simultaneously  modeled with the photo-ionization code \texttt{CLOUDY}, allowing us to infer physical conditions in the ionized and neutral phases of the ISM.}
   {Starburst-dominated galaxies exhibit systematic radial trends, including a decreasing ionization parameter with galactocentric distance, an approximately constant hydrogen density, uniformly young stellar populations, and flat metallicity profiles. The ionization parameter gradients derived from MIR-only diagnostics indicate an increasing trend with radius, while diagnostics incorporating FIR lines instead reveal a declining radial behavior. This discrepancy highlights the critical role of FIR tracers, especially [C\textsc{ii}]$_{158, \rm PDR}$, in constraining the ionization structure of the ISM. Diagnostic diagrams combining MIR and FIR tracers show that in the strongest AGN-dominated system of our sample, the nucleus and inner extended regions (projected radial distances $\lesssim$5~kpc) occupy areas in the parameter space with elevated [Ne~III]/[Ne~II] and suppressed [C~II]/[O~I] ratios, while only the outermost regions ($\gtrsim$~6~kpc) show values typical of star-forming regions. This radial variation suggests that the AGN might affect the excitation conditions of the ionized and neutral gas out to projected galactocentric distances of approximately 5~kpc.}
   {Our results highlight the importance of combining MIR and FIR diagnostics to recover and quantify physically meaningful radial trends in luminous dusty galaxies, and they show that galaxy-integrated measurements can mask substantial spatial variations. In combination with the results of paper I, this comprehensive work establishes a large-scale benchmark for studies of gas and dust grain properties in starbursts and AGN-hosting galaxies with a high spatial resolution obtained with facilities such as the James Webb Space Telescope (JWST) and the Atacama Large Millimeter/submillimeter Array (ALMA).}


   \maketitle
%
\section{Introduction}
Infrared (IR) spectroscopy, particularly at mid- and far-infrared wavelengths (MIR and FIR, respectively), plays a central role in the study of the interstellar medium (ISM) in galaxies by providing access to gas and dust components that are heavily obscured at optical and ultraviolet wavelengths. A significant fraction of the physical processes associated with star formation and nuclear activity in massive galaxies takes place in dust-embedded environments, where extinction substantially limits the diagnostic power of traditional tracers. Infrared observations, by contrast, are only weakly affected by dust attenuation and therefore offer direct probes of the physical conditions in obscured regions of the ISM \citep{Spinoglio92}. IR fine-structure emission lines enable a powerful set of diagnostics with which we can investigate the multiphase structure of the ISM. Their line ratios are well known to be sensitive to key physical parameters such as the hardness of the radiation field, the ionization parameter, the gas density, and the metallicity \citep[e.g.][]{DeCarli25}. MIR lines trace photoionized gas in \ion{H}{II} regions and are widely used to constrain the properties of the ionizing stellar population, while FIR lines dominate the cooling of neutral gas in photodissociation regions (PDRs), which are the interface zones between ionized and molecular gas \citep{Tielens85, Wolfire90}. 

During the past two decades, the Infrared Spectrograph (IRS) on board the Spitzer Space Telescope (\textit{Spitzer} hereafter; \citet{Werner04}) enabled uniform spectroscopic observations of large and diverse galaxy samples. Measurements of fine-structure emission lines such as [Ne\,\textsc{ii}]$_{13}$ and [Ne\,\textsc{iii}]$_{15}$ provided constraints on the properties of the ionizing stellar population, the hardness of the radiation field \citep{Ho07}, and the relative contributions of star formation and activity from active galactic nuclei (AGNs) \citep{Tommasin08, Tommasin10}, while density-sensitive ratios including [S\,\textsc{iii}]$_{33}$/[S\,\textsc{iii}]$_{19}$ allowed us to obtain extinction-independent estimates of electron densities in ionized gas.
At longer wavelengths, the \textit{Herschel Space Observatory} (\textit{Herschel} hereafter; \citet{Pilbratt10}) extended these diagnostics into the FIR with improved sensitivity. Spectroscopy with the Photodetector Array Camera and Spectrometer (PACS) and the Spectral and Photometric Imaging Receiver (SPIRE) on board Herschel provided access to the dominant cooling lines of the neutral ISM, [C\textsc{ii}]$_{158}$ and [O\textsc{i}]$_{63}$, which trace the cold neutral gas \citep{Tielens85, Spinoglio15}. Large Herschel surveys revealed variations in the FIR line emission with global galaxy properties, demonstrating the sensitivity of these diagnostics to changes in ionization parameter, gas density, and dust--gas coupling \citep{Gracia11, Farrah13, Tanio13, Tanio14, Rosenberg15}. 

While many IR spectroscopic studies have focused on nuclear and/or galaxy-integrated measurements, spatially resolved observations have shown that the ISM properties change significantly on kiloparsec scales. MIR spectral mapping with \textit{Spitzer} revealed extended ionized gas and star formation in disks and circumnuclear regions, traced by low-ionization lines, and spatial variations in the radiation hardness \citep{Armus04, Armus06, Pereira-Santaella10, Tanio11}. FIR observations further showed that several FIR lines exhibit deficits with increasing dust temperature, most notably [C\textsc{ii}]$_{158}$, [N\textsc{ii}]$_{122}$, and [N\textsc{ii}]$_{205}$. The deficit in the [O\textsc{i}]$_{63}$ line is weaker, whereas [O\textsc{iii}]$_{88}$ displays no systematic deficit but increased scatter at higher dust temperatures \citep{Tanio17}.
Integrated analyses combining \textit{Spitzer} and \textit{Herschel} observations have demonstrated the diagnostic power of this approach in disentangling the relative roles of star formation, AGN activity, and shocks in powering the IR emission of luminous infrared galaxies (LIRGs, \citet{Fernandez-Ontiveros16, Herrera18}). However, the majority of these combined \textit{Spitzer}+\textit{Herschel} studies relied on galaxy-integrated measurements for entire galaxies. Spatially resolved analyses that combine these datasets have remained challenging because their sensitivities and spatial resolutions are different. 

This work is the second of a two-part series of papers that investigates the ISM in a subsample of the Great
Observatories All-sky LIRG Survey (GOALS) galaxies \citep{Armus09}. The first paper (Paper~I; \citet{Lofaro26}) focused on the resolved measurements of polycyclic aromatic hydrocarbons (PAHs), while this work examines the spatial structure and physical properties of the ISM in the subsample. Our analysis targets regions located well beyond the nuclear zones and extends to galactocentric distances that exceed the spatial coverage typically probed by a single pointing of any James Webb Space Telescope (JWST) integral-field unit (IFU). We focus on how AGN activity and star formation shape the ionized and neutral gas in different galactic environments. By combining spatially resolved MIR maps from \textit{Spitzer} with IFU measurements from \textit{Herschel}, we present a joint spatially resolved analysis of LIRGs on kiloparsec scales. 

The paper is organized as follows. Sect.~\ref{Goals sample} describes the sample selection and the observations used in this study. In Sect.~\ref{section3} we present the data-processing steps, including spectral extraction and emission-line-fitting procedures. Sect.~\ref{models} outlines the \texttt{CLOUDY} photoionization models employed to interpret the data. The main results and their interpretation are discussed in Sect.~\ref{results} and Sect.~\ref{radial profile}. Finally, we summarize our conclusions in Sect.~\ref{conclusions}.

Throughout this paper, we assume a flat $\Lambda$CDM cosmology with $H_0 = 67.4$ km s$^{-1}$ Mpc$^{-1}$, $\Omega_{\rm m} = 0.315$, and $\Omega_{\Lambda} = 0.685$ \citep{Plank20}.

\section{Sample and observations}
\label{Goals sample}
\color{black}

\subsection{The GOALS sample}
This work is based on a subsample of 67 galaxies drawn from the GOALS survey \citep{Armus09}, a multiwavelength program targeting (U)LIRGs selected from the Infrared Astronomical Satellite (IRAS) Revised Bright Galaxy Sample. The sources have available spectral mapping observations from \textit{Spitzer}/IRS and \textit{Herschel}/PACS, span a wide range of infrared luminosities, nuclear classifications, and interaction stages, and are broadly representative of the full GOALS population. A detailed description of the parent sample, selection criteria, and global properties is provided in Paper~I. 

The GOALS survey has enabled a series of key emission line studies in multiple wavelength regimes. Using \textit{Spitzer}/IRS high-resolution nuclear spectra, \citet{Inami13} presented a homogeneous MIR fine-structure line catalog for the full GOALS sample. By comparing line ratios with photoionization and shock models, they constrained typical starburst ages, ionization parameters, metallicities, and electron densities in GOALS nuclei and identified a subset of sources with signatures consistent with shock excitation or stratified ionized gas. In the FIR regime, \citet{Tanio13, Tanio14, Tanio17} used \textit{Herschel}/PACS spectroscopy to study the behavior of [C\,II]$_{158}$, [O\,I]$_{63}$, [O\,III]$_{88}$, and [N\,II]$_{122}$ emission in GOALS galaxies, establishing the existence of systematic FIR line deficits and demonstrating their strong dependence on IR surface brightness, dust temperature, and source compactness. These studies further revealed that the [C\,II]$_{158}$ emission in LIRGs is often spatially extended, highlighting the importance of separating nuclear and extra-nuclear contributions when interpreting far-IR diagnostics.
At optical wavelengths, integral-field spectroscopy of GOALS galaxies by \citet{Rich14} showed that shocks play a major role in shaping the emission line properties of merging LIRGs, particularly in advanced interaction stages, and that shock excitation can contribute significantly to the observed optical line ratios and kinematics. Together, these works have established GOALS as the local benchmark for emission line studies of luminous infrared galaxies. However, most previous analyses either focused on nuclear or globally integrated properties or on individual wavelength regimes. In this paper, we build upon these foundations by combining spatially resolved MIR and FIR emission line measurements to investigate the physical conditions of the ISM in local LIRGs.

\subsection{Spitzer/IRS observations}
\label{CUBISM}
The galaxies were observed with the \textit{Spitzer}/IRS in spectral mapping mode using the low-resolution Short-Low (SL) and Long-Low (LL) modules. The data were retrieved from the \textit{Spitzer} Heritage Archive\footnote{\href{https://sha.ipac.caltech.edu/applications/Spitzer/SHA/}{https://sha.ipac.caltech.edu/applications/Spitzer/SHA/}} (PID~30323; PI~L.~Armus), together with additional archival observations.
The construction of the IRS spectral cubes and the data reduction procedures followed those described in paper~I \citep{Lofaro26}, to which we refer for more details. The \textit{Spitzer}/IRS fine-structure emission lines analyzed in this work included
[\ion{Ar}{ii}]$_{7\mu m}$, [\ion{Ar}{iii}]$_{9\mu m}$, [\ion{S}{iv}]$_{10\mu m}$,
[\ion{Ne}{ii}]$_{13\mu m}$, [\ion{Ne}{iii}]$_{15\mu m}$,
[\ion{S}{iii}]$_{19\mu m}$, [\ion{O}{iv}]$_{25\mu m}$, and [\ion{S}{iii}]$_{33\mu m}$.

\subsection{Herschel/PACS observations}
\label{Herschel observations}
We used FIR spectroscopic observations obtained with the PACS Integral Field Spectrometer on board \textit{Herschel}, drawn primarily from large Herschel programs (PI: L.~Armus) and supplemented by archival datasets. These observations provide spectral maps of the [O\textsc{i}]$_{63}$, [O\textsc{iii}]$_{88}$, and [C\textsc{ii}]$_{158}$ emission lines. Further details of the data properties are described in \citet{Tanio13} and in paper~I \citep{Lofaro26}. 

\color{black}
\section{Spectral extraction and emission line measurements}
\label{section3}
\subsection{Spectral extraction analysis}
\label{spectral extraction}
Spatially resolved measurements were obtained by applying a grid of square apertures across the full extent of each source, allowing us to probe local variations in the ISM. A detailed description of the grid construction, aperture definition, and extraction method is provided in paper~I.

Briefly, the \textit{Spitzer} extraction grid was defined such that each aperture corresponds to a physical size of 2~kpc on a side, after rotating all images to a common orientation in which north is up, and east is left (\textit{Spitzer grid} configuration). For the \textit{Herschel}/PACS data, the line fluxes were extracted using square apertures of $7.5\arcsec$ per side, which is slightly smaller than the original PACS spaxels. The same grid geometry was then applied to the \textit{Spitzer}/IRS cubes to ensure a consistent spatial comparison between the two datasets (\textit{Herschel grid} configuration; see Fig.~2 in paper~I). 

\subsection{Spectral fitting}
The MIR spectra obtained with Spitzer/IRS were analyzed using a modified implementation of the code continuum and feature extraction (CAFE\footnote{\href{https://github.com/GOALS-survey/CAFE}{https://github.com/GOALS-survey/CAFE}}; \citet{Marshall07, CAFE25}, see Sect.~\ref{cafe_fits} of the appendix). Dust attenuation corrections were applied to the derived line fluxes, assuming a mixed geometry for PAHs and a foreground screen geometry for the warm and hot dust continua. 

For the FIR data from Herschel/PACS, the emission line fluxes were measured by fitting Gaussian profiles to the lines after subtracting a locally defined linear continuum. Absolute calibration uncertainties of up to $\mathord{\sim}$11--12\%\footnote{\href{http://herschel.esac.esa.int/Docs/PACS/html/ch04s10.html}{http://herschel.esac.esa.int/Docs/PACS/html/ch04s10.html}} were incorporated by adding them in quadrature to the statistical errors. Further details of the fitting strategy and uncertainty treatment are presented in paper~I.

\section{Photoionization models}
\label{models}
We compared our fine-structure emission line measurements with a series of photoionization model grids produced using the spectral synthesis code \texttt{CLOUDY}\footnote{\href{https://trac.nublado.org/}{https://trac.nublado.org/}} (version C23; \citet{Cloudy23}), in order to constrain the physical conditions of the ISM in our sample of galaxies. To compute the photoionization models of starburst galaxies, a spectral energy distribution (SED) is required as the ionizing radiation field. We adopted SEDs from the PopStar stellar population synthesis models \citep{Popstar} (see Appendix \ref{popstar}), which are based on the Padova evolutionary tracks and assume a Kroupa initial mass function \citep{Kroupa02}, with stellar masses ranging from 0.15 to 100 $M_{\odot}$. As LIRGs are typically characterized by merger-driven bursts of star formation, we used an instantaneous star formation history to more accurately reflect the bursty nature of their stellar populations. We explored metallicities of $Z=Z_{\odot}$ and $Z=2.5Z_{\odot}$ \citep{Inami13, Rich12}, corresponding to the only solar and super-solar values available in PopStar. We used models with constant pressure \citep{Abel05, Fernandez-Ontiveros16}, a hydrogen column density of $N_{\rm H}=5\times10^{22}~{\rm cm^{-2}}$, stellar ages between 1 and 5 Myr in steps of 0.5 Myr, hydrogen densities in the range $1 \leq \log(n_{\rm H}/{\rm cm^{-3}}) \leq 4$ in steps of 0.5 dex, and ionization parameters $-4.5 \leq \log U \leq -2.5$ in steps of 0.25 dex \citep{Spinoglio15, Fernandez-Ontiveros16, Pereira17}. The ionization parameter was defined as follows:
\begin{equation}
    U \equiv \frac{Q(H)}{4\pi r_0^2 n(H)c} \equiv \frac{\Phi(H)}{n(H)c},
\end{equation}
\noindent where Q(H) is the number of hydrogen-ionizing photons, r$_0$ is the distance between the ionizing source and the cloud, n$_H$ is the hydrogen density of the cloud, \textit{c} is the speed of light, and $\Phi(H)$ is the flux of hydrogen-ionizing photons. The choice of the column density ensured that we reached the PDRs, as indicated by the fact that the outermost slab of the cloud contains about 92$\%$ neutral hydrogen and exhibits a temperature of $\mathord{\sim}$60--70 K. We also tested older stellar populations of 10 Myr, but they failed to reproduce the observed line ratios. This agrees with previous GOALS studies, which showed that young stellar ages of 1--4.5 Myr are required to match the MIR fine-structure diagnostics \citep{Inami13, Tanio17}. 

To model the AGN-dominated galaxies, we adopted a pure AGN SED composed of two components: a big bump, described by a power law rising to a peak temperature $T_{\rm BB}$ with a low-energy slope $\alpha_{\rm UV}$ and infrared and high-energy exponential cutoffs and an X-ray continuum from 1.36~eV to 100~keV with a slope $\alpha_{\rm X}$.
The normalization between these two components is set by the X-ray to UV ratio, $\alpha_{\rm OX}$, which is defined as
\begin{equation}
    \alpha_{OX} = \frac{\log(\rm L_{2keV}/\rm L_{2500\mathring{A}})}{2.605}.
\end{equation}
We used a constant density \citep{Fernandez-Ontiveros16, Fernandez-Aranda24} and the default values $\alpha_{\rm OX}=-1.4$ and $T_{\rm BB}=10^{6}$~K to avoid introducing degeneracies in the fitting procedure because $T_{\rm BB}$ is strongly degenerate with $\alpha_{\rm OX}$ and the ionization parameter $U$. 
All models were computed assuming the default grain and PAH size distribution described in \cite{Abel08}. In addition to these dust properties, we also accounted for observed chemical abundance patterns in starburst galaxies. Specifically, sulfur appears to be underabundant in the gas phase of high-metallicity starburst galaxies compared to solar values. This sulfur underabundance is interpreted as the result of sulfur depletion onto dust grains and not as an intrinsic nucleosynthetic effect. Following the results of \cite{Verma03} and \cite{Fernandez-Ontiveros16}, we adopted a sulfur abundance equal to two-thirds of the solar value in our models to reflect this depletion. 
We note that the high-ionization lines ([S\textsc{iv}]$_{10}$ and [O\textsc{iv}]$_{26}$; 34.86~eV and 54.93~eV, respectively) tend to be underpredicted by photoionization models, including \texttt{CLOUDY}, suggesting that additional excitation mechanisms might be required. Consequently, we excluded diagnostic diagrams involving these transitions from our \texttt{CLOUDY} grid analysis. To explore potential alternatives, we tested alternative stellar spectral energy distributions using the binary population and spectral synthesis (BPASS; \citet{BPASS}) models, which incorporate  binary stellar evolution. However, this approach did not yield significant 
improvements in reproducing the high-ionization line emission. This modeling  challenge is well documented in the literature: \cite{Inami13} found that even shock models from \cite{Allen08} were only able to reproduce [S\textsc{iv}]$_{10}$ emission in 10 out of 200 GOALS galaxies.  While the persistent difficulty in modeling high-ionization transitions through photoionization alone represents a fundamental limitation that requires further investigation, this effort is beyond the scope of this paper. 

\section{Results}
\label{results} 
In the following sections, we investigate multiple line ratios that trace specific physical conditions of the ionized and neutral gas in galaxies on spatially resolved scales. Each diagnostic diagram is compared with the photoionization model grids described in Sect.~\ref{models}. The model grids relevant to the diagnostics discussed here are shown in this section, while additional model grids and line-ratio diagnostics are presented in Appendix~\ref{Additional models Appendix}.

\subsection{\textit{The resolved [S\textsc{iii}]$_{33}$/[S\textsc{iii}]$_{19}$ line ratio}}
\label{SIII}

\begin{figure*}[t]
    \sidecaption
    \includegraphics[width=12cm]{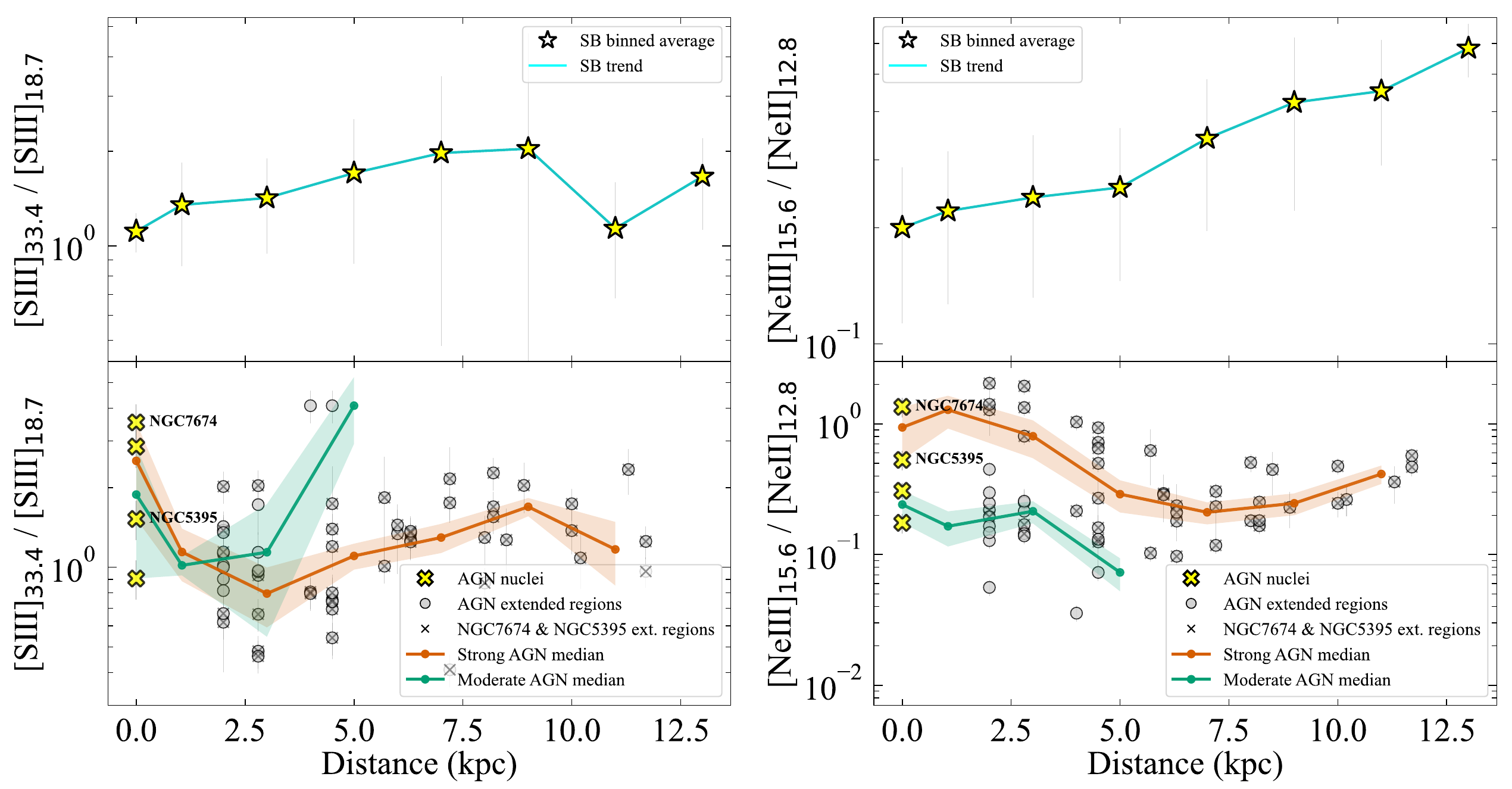} 
    \caption{Radial distributions of [S\textsc{iii}]$_{33}$/[S\textsc{iii}]$_{19}$ (left) and [Ne\textsc{iii}]$_{15}$/[Ne\textsc{ii}]$_{13}$ (right) as a function of projected galactocentric distance. The upper panels show the starburst-dominated regions ($f_{\mathrm{AGN}}{<}0.4$), while the lower panels show the AGN-hosting systems ($f_{\mathrm{AGN}}{\geq}0.4$). For the starburst-dominated galaxies, yellow stars indicate the binned average values, with the cyan line showing their radial trend. For the AGN-hosting galaxies, yellow crosses mark the nuclei, and gray circles indicate the extended regions. The extended regions of NGC~7674 and NGC~5395 are further highlighted with an overlaid cross. The orange and green lines show the median radial trends for the strongest and moderate AGN systems, respectively, with shaded regions indicating the error on the median.}
    \label{fig:Ne_SIII_radial_trend}
\end{figure*}

\begin{figure*}[] 
    \centering
    \includegraphics[width=\textwidth]{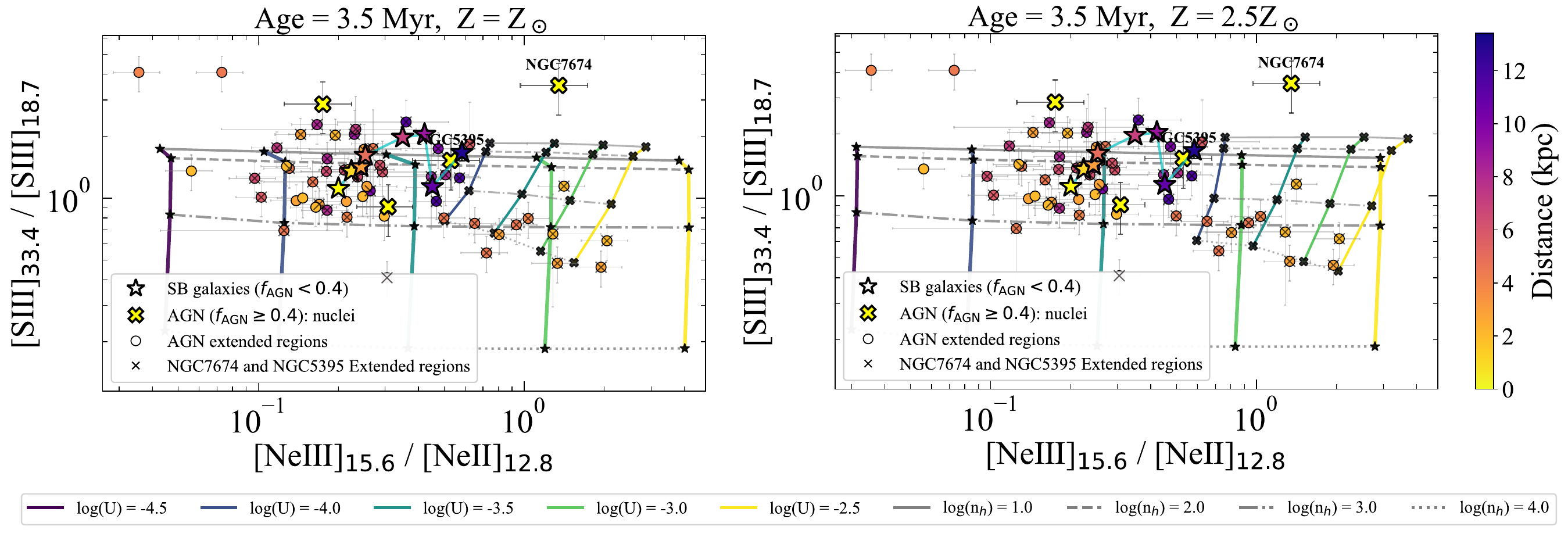} 
    \caption{Diagnostic diagram of [S\textsc{iii}]$_{33}$/[S\textsc{iii}]$_{19}$ vs. [Ne\textsc{iii}]$_{15}$/[Ne\textsc{ii}]$_{13}$, tracing the electron density and radiation field hardness, respectively. The symbols follow the same convention as in Fig.~\ref{fig:Ne_SIII_radial_trend}, but are color-coded according to their galactocentric distance. The photoionization model grids computed for stellar ages of 3.5 Myr and metallicities of $Z{=}Z_{\odot}$ (\textit{left panel}) and $Z{=}2.5Z_{\odot}$ (\textit{right panel}) are overlaid. Additional grids for stellar ages of 1 and 5 Myr are shown in Appendix~\ref{Additional models Appendix}. The starburst and AGN model grids are shown with stars and crosses, respectively. Each curve corresponds to a fixed hydrogen density, with $\log(n_{\mathrm{H}})$ increasing from top to bottom, while the ionization parameter, $\log(U)$, increases from left to right.}
    \label{fig:SIII}
\end{figure*}

The MIR [S\textsc{iii}]$_{33}$/[S\textsc{iii}]$_{19}$ ratio is widely used as an electron-density probe for H\textsc{ii} region gas, and it typically traces low to moderate densities in starburst galaxies and AGN hosts. The [S\textsc{iii}]$_{33}$/[S\textsc{iii}]$_{19}$ ratio decreases with increasing electron density because the critical densities of the two transitions are different, approximately $\mathord{\sim} 2\times10^{3}$ cm$^{-3}$ for [S\textsc{iii}]$_{33}$ and $\mathord{\sim} 4\times10^{4}$ cm$^{-3}$ for [S\textsc{iii}]$_{19}$. Thus, this ratio probes densities in the range of $\mathord{\sim} 10^{2}- 10^{5}$ cm$^{-3}$ \citep{Martin02, Dale06}. The GOALS LIRGs show a median [S\textsc{iii}]$_{33}$/[S\textsc{iii}]$_{19}$ of ${\thickapprox}$ 1.61, corresponding to a typical electron density of  n$_e$ ${\thickapprox}$ 300 cm$^{-3}$ \citep{Inami13}. \\
The diagnostic plot of [S\textsc{iii}]$_{33}$/ [S\textsc{iii}]$_{19}$ versus [Ne\textsc{iii}]$_{15}$/ [Ne\textsc{ii}]$_{13}$ provides an almost orthogonal characterization of the electron density and radiation field hardness. The spatially resolved mapping of LIRGs shows that the variations in excitation, traced by [Ne\textsc{iii}]$_{15}$/ [Ne\textsc{ii}]$_{13}$, are typically stronger than changes in density across galactic centers and disks: the nuclei often display lower [Ne\textsc{iii}]$_{15}$/[Ne\textsc{ii}]$_{13}$ than the surrounding star-forming regions \citep{Pereira-Santaella10}. Taken together with the nuclear [S\textsc{iii}]$_{33}$/[S\textsc{iii}]$_{19}$ measurements from large \textit{Spitzer} samples, the overall picture in a [S\textsc{iii}]$_{33}$/ [S\textsc{iii}]$_{19}$ versus [Ne\textsc{iii}]$_{15}$/[Ne\textsc{ii}]$_{13}$ diagram is that starbursts and AGNs are primarily separated along the horizontal hardness axis, while extended regions cluster around [S\textsc{iii}]$_{33}$/ [S\textsc{iii}]$_{19}$ values corresponding to $\rm n_e$ $\leq 10^{3}$ cm$^{-3}$. 

Figure~\ref{fig:Ne_SIII_radial_trend} shows the [S\textsc{iii}]$_{33}$/[S\textsc{iii}]$_{19}$ and [Ne\textsc{iii}]$_{15}$/[Ne\textsc{ii}]$_{13}$ radial trends for the nuclear and resolved regions of our sample. In Fig.~\ref{fig:SIII} and Fig.~\ref{fig:SIII_appendix} we compare the observational data with \texttt{CLOUDY} photoionization model grids (described in Sect.~\ref{models}) computed for different nuclear power sources, stellar ages, metallicities, ionization parameters, and gas densities (see the caption of Fig.~\ref{fig:SIII} for more details). \\

\indent Starburst galaxies. In the upper panel of Fig.~\ref{fig:Ne_SIII_radial_trend}, starburst regions exhibit an overall increase in the [S\textsc{iii}]$_{33}$/[S\textsc{iii}]$_{19}$ ratio from the nuclei out to galactocentric distances of $\sim$8~kpc, with the ratio increasing from $\mathord{\sim}$1.2 to $\mathord{\sim}$2.2, while at larger radii, the ratio exhibits increased scatter. This behavior might reflect a broader range of radiation field conditions, as suggested by the increase in [Ne\textsc{iii}]/[Ne\textsc{ii}] with galactocentric distance from ${\sim}$0.2 to ${\sim}$0.6. The hydrogen density inferred from the models spans $1\lesssim\log(n_{\rm H}/\mathrm{cm}^{-3})\lesssim3$ (see Fig.~\ref{fig:SIII}), with a median value of $\log(n_{\rm H}/\mathrm{cm}^{-3}){\sim}2.3$, encompassing nuclear and extended regions. In contrast, the ionization parameter appears to differ slightly when we compare nuclear and extended regions: at $\rm Z_{\odot}$ and 3.5~Myr, the nuclear regions are concentrated around $-4{\lesssim}\log U{\lesssim}-3.5$, whereas extended regions cluster around $-4{\lesssim}\log U{\lesssim}-3$. This is consistent with a picture in which the ISM becomes progressively more complex and spatially variable away from the nucleus (see also \citet{Dale06}).
Overall our results suggest that moderately hard radiation fields produced by young massive stars are sufficient to reproduce the observed [S\textsc{iii}]$_{33}$/ [S\textsc{iii}]$_{19}$ and [Ne\textsc{iii}]$_{15}$/[Ne\textsc{ii}]$_{13}$ ratios in typical star-forming environments. This agrees with previous galaxy-integrated studies of starburst galaxies \citep[e.g.,][]{Fernandez-Ontiveros16}. \\

\indent AGN-dominated galaxies.
The AGN-dominated nuclei exhibit a broader diversity of physical conditions. In the lower panels of Fig.~\ref{fig:Ne_SIII_radial_trend}, the nucleus of NGC~7674, one of the galaxies with the highest AGN fractions in our sample, displays the highest [Ne\textsc{iii}]$_{15}$/[Ne\textsc{ii}]$_{13}$ ratio (${\sim}1.4$) among the AGN-dominated nuclei, together with a high [S\textsc{iii}]$_{33}$/[S\textsc{iii}]$_{19}$ ratio of ${\sim}3.5$. NGC~7469, which has a moderate AGN fraction, also exhibits an elevated [S\textsc{iii}]$_{33}$/[S\textsc{iii}]$_{19}$ ratio of ${\sim}2.9$, despite its substantially lower [Ne\textsc{iii}]$_{15}$/[Ne\textsc{ii}]$_{13}$ ratio of ${\sim}0.18$. The central values of these two sulfur line ratios exceed the theoretical low-density limit of ${\sim}2$--$2.5$ and therefore cannot be interpreted quantitatively as evidence of very low electron densities. Their measurement uncertainties must also be taken into account. These elevated ratios might reflect uncertainties in measuring the relatively weak 18.7~$\mu$m line against the strong continuum, or residual differential extinction that preferentially attenuates the 18.7~$\mu$m emission relative to the 33.5~$\mu$m line. By comparison, NGC~5395, the other galaxy with a high AGN fraction, and NGC~5990, which has a moderate AGN fraction, show lower nuclear [S\textsc{iii}]$_{33}$/[S\textsc{iii}]$_{19}$ ratios of ${\sim}1.5$ and ${\sim}0.9$, respectively, with corresponding [Ne\textsc{iii}]$_{15}$/[Ne\textsc{ii}]$_{13}$ ratios of ${\sim}0.53$ and ${\sim}0.31$. The extended regions of the AGN-dominated galaxies span a broader range of observed ratios, most notably in NGC~7674. In this system, the emission at galactocentric distances of $\sim$1--5~kpc shows [Ne\textsc{iii}]$_{15}$/[Ne\textsc{ii}]$_{13}$ ratios that are comparable to that of the nucleus, but significantly lower [S\textsc{iii}]$_{33}$/[S\textsc{iii}]$_{19}$ values, by $\mathord{\sim}80\%$.
The median radial trends shown in the bottom panels of Fig.~\ref{fig:Ne_SIII_radial_trend} further illustrate the different behavior of the strong- and moderate-AGN subsamples. The strong-AGN median remains systematically higher in [Ne\textsc{iii}]$_{15}$/[Ne\textsc{ii}]$_{13}$ than the moderate-AGN median over most of the radial range, indicating harder radiation fields. The [S\textsc{iii}]$_{33}$/[S\textsc{iii}]$_{19}$ ratio shows a more variable behavior, with the moderate-AGN median increasing at intermediate radii, suggesting possible changes in the density structure of the ionized gas. More generally, the moderate-AGN median lies closer to the locus of the starburst-dominated regions, particularly in [Ne\textsc{iii}]$_{15}$/[Ne\textsc{ii}]$_{13}$, suggesting that the extended regions of these systems are less strongly affected by the AGN. This comparison should be interpreted cautiously given the limited number of AGN-hosting galaxies, but it suggests that the strongest AGN might affect the ionized-gas conditions over larger spatial scales than the moderate-AGN systems.
The model grids shown in Fig.~\ref{fig:SIII} and Fig.~\ref{fig:SIII_appendix} provide further insight into these trends. The combination of high [Ne\textsc{iii}]$_{15}$/[Ne\textsc{ii}]$_{13}$ and [S\textsc{iii}]$_{33}$/[S\textsc{iii}]$_{19}$ ratios observed in the nucleus of NGC~7674 indicates a hard radiation field ($\log U{\sim}-3$) acting on relatively low-density gas. However, the hydrogen density of its nuclear region and the nuclear region of NGC~7469 cannot be robustly constrained because the \texttt{CLOUDY} models underpredict the observed [S\textsc{iii}]$_{33}$/[S\textsc{iii}]$_{19}$ ratio. The lower [S\textsc{iii}]$_{33}$/[S\textsc{iii}]$_{19}$ values observed in the nuclei of NGC~5990 and NGC~5395 instead indicate moderately higher densities, with $2\lesssim\log(n_{\rm H}/\mathrm{cm}^{-3})\lesssim3$, while their [Ne\textsc{iii}]$_{15}$/[Ne\textsc{ii}]$_{13}$ ratios indicate moderately hard radiation fields, with $\log U{\sim}-4$ to ${\sim}-3$. In the extended regions of NGC~7674, the substantially lower [S\textsc{iii}]$_{33}$/[S\textsc{iii}]$_{19}$ ratios correspond to average gas densities reaching ${\sim}10^{3}$\,cm$^{-3}$. These densities are higher than those typically observed in starburst nuclei and extended star-forming regions. Such elevated densities are commonly observed in AGN-dominated systems and are consistent with those inferred for Seyfert narrow-line regions using MIR fine-structure diagnostics, for which electron densities of about $n_e{\sim}10^{3}$--$10^{4}$~cm$^{-3}$ have been reported (e.g., \citealt{Tommasin08}). Finally, the starburst model grids provide good coverage of the observed ratios in NGC~5395, NGC~5990, NGC~7469, and NGC~7674, for which the AGN model grid is also shown.

\subsection{\textit{The resolved [C\textsc{ii}]$_{158}$/[O\textsc{i}]$_{63}$ line ratio}}
\label{CII/OI}

\begin{figure*}[t]
    \sidecaption
    \includegraphics[width=12cm]{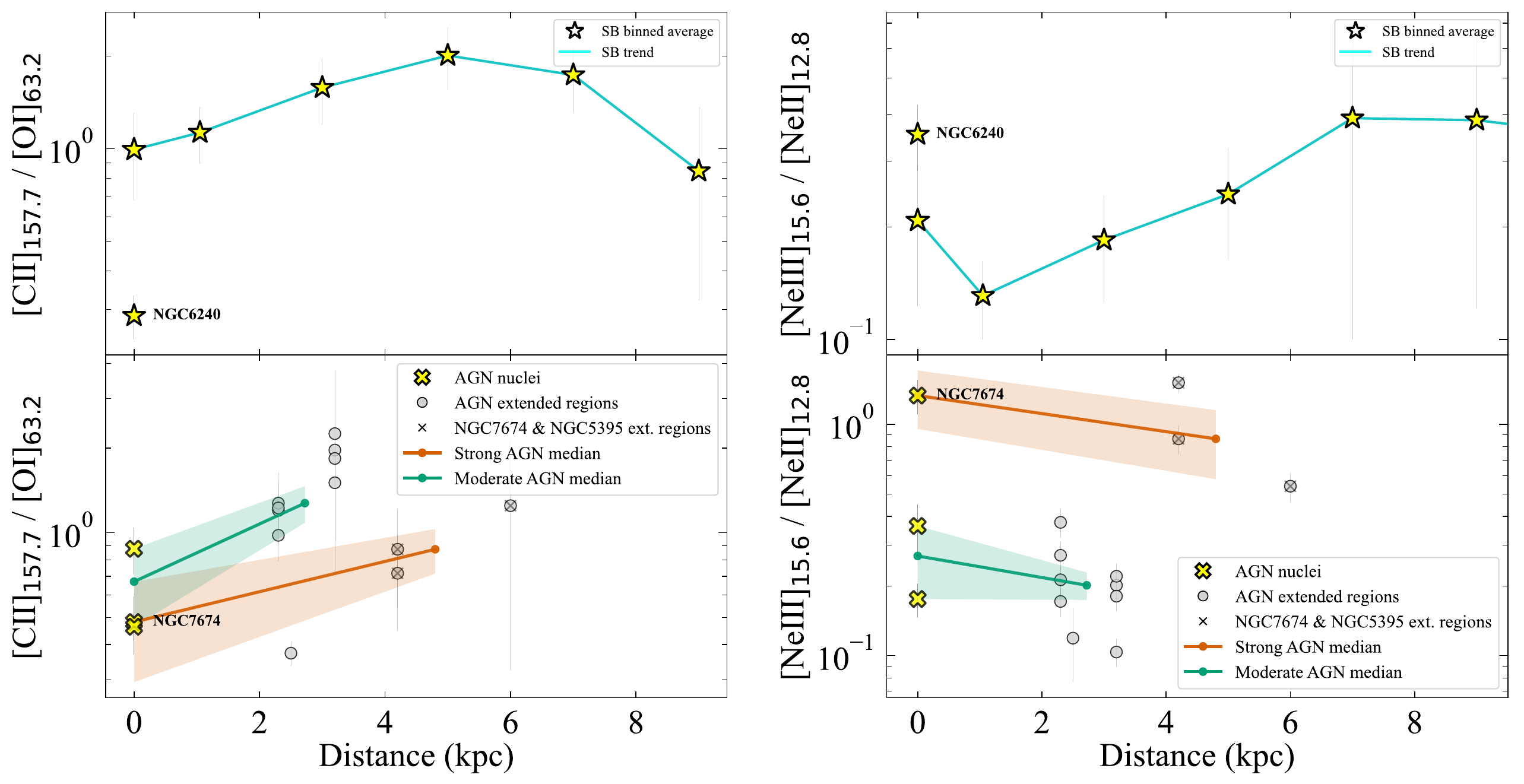}
    \caption{Radial distributions of [C\textsc{ii}]$_{158}$/[O\textsc{i}]$_{63}$ (left) and [Ne\textsc{iii}]$_{15}$/[Ne\textsc{ii}]$_{13}$ (right) as a function of projected galactocentric distance in the \textit{Herschel grid} configuration. The symbols follow the same convention as in Figure \ref{fig:Ne_SIII_radial_trend}.}
    \label{fig:CII_OI_radial_trend}
\end{figure*}

\begin{figure*}[] 
    \centering
    \includegraphics[width=\textwidth]{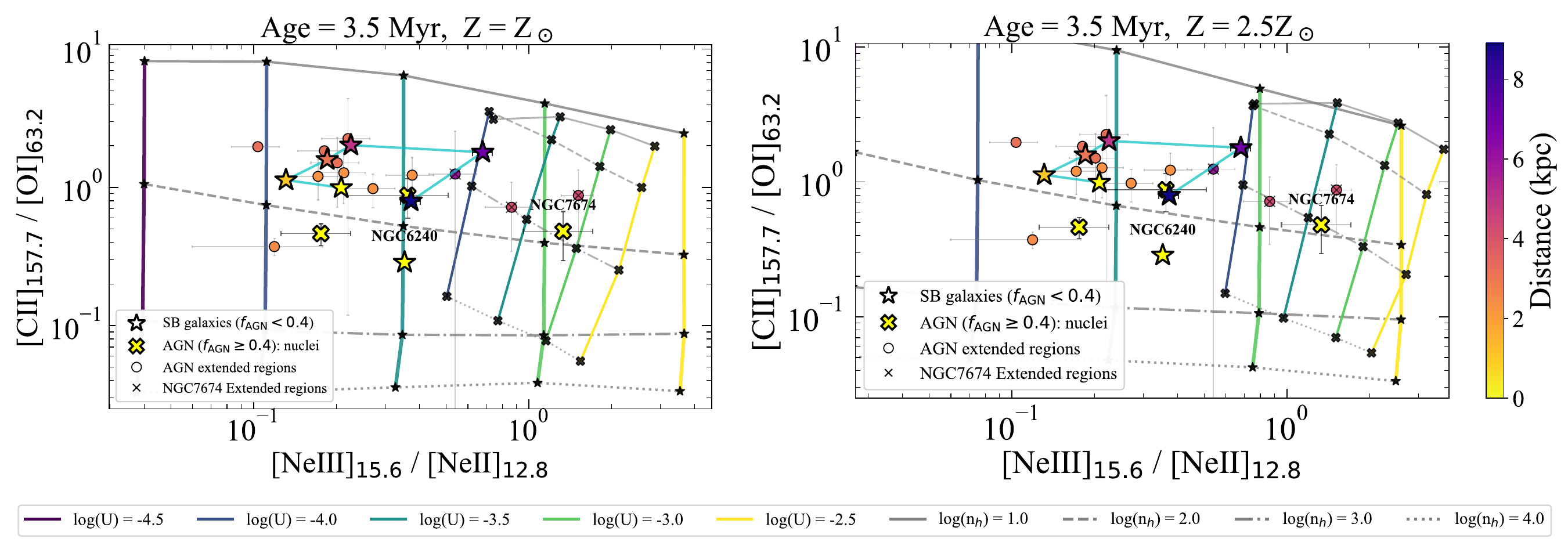}
    
    \caption{Diagnostic diagram of log([C\textsc{ii}]$_{158}$/[O\textsc{i}]$_{63}$) vs. log([Ne\textsc{iii}]$_{15}$/[Ne\textsc{ii}]$_{13}$), probing the neutral gas cooling conditions and radiation field hardness, respectively. The symbols follow the same convention as in Figure \ref{fig:SIII}. Additional grids for stellar ages of 1 and 5 Myr are shown in Appendix~\ref{Additional models Appendix}.}
    \label{fig:CII_OI}
\end{figure*}

The PDRs are predominantly neutral gas transition layers between H\textsc{ii} regions and molecular clouds, where far-ultraviolet (FUV) photons regulate the chemical, thermal, and ionization balance of the gas. These regions are heated primarily by the photoelectric effect on dust grains, and they cool efficiently through IR fine-structure lines, most notably, [C\textsc{ii}]$_{158}$ and [O\textsc{i}]$_{63}$. Together, these lines account for the bulk of the gas cooling in PDRs \citep{Rosenberg15}, with [C\textsc{ii}]$_{158}$ dominating at moderate temperatures and densities (n$_{crit}\mathord{\sim}$ 2$\times10^{3}$ cm$^{-3}$),  
and [OI]$_{63}$ becoming more significant in denser (n$_{crit}\mathord{\sim}$ 9$\times10^{5}$ cm$^{-3}$), warmer gas (T $\geq$ 300 K; see Fig.~6 in \citet{Tanio17}). Their relative strengths are widely used to probe the temperature (\citet{Ho15}) and the incident FUV field strength (high [C\textsc{ii}]$_{158}$/[O\textsc{i}]$_{63}$ ratios signal extended low-density PDRs, while low values trace compact and dense neutral gas; e.g., \citet{Hollenbach99, Kaufman99, Malhotra01, Wolfire22}).  
The diagnostic diagram of [C\textsc{ii}]$_{158}$/[O\textsc{i}]$_{63}$ versus [Ne\textsc{iii}]$_{15}$/[Ne\textsc{ii}]$_{13}$ combines information about the cooling of the neutral ISM with the intensity of the ionizing radiation field. \\

\indent Starburst galaxies.
Based on the \textit{Herschel} grid configuration, the [Ne\textsc{iii}]$_{15}$/[Ne\textsc{ii}]$_{13}$ ratio at $\sim$5\,kpc marks a change in the radial behavior of the [C\textsc{ii}]$_{158}$/[O\textsc{i}]$_{63}$ ratio. In the upper panel of Fig.~\ref{fig:CII_OI_radial_trend}, within the inner $\sim$5\,kpc, [C\textsc{ii}]$_{158}$/[O\textsc{i}]$_{63}$ shows a mild increase with galactocentric distance, rising from $\mathord{\sim}1$ to $\mathord{\sim}2$. This is consistent with neutral gas that is more diffuse and characterized by lower average densities. Beyond $\sim$6\,kpc, the ratio gradually decreases toward values that are comparable to those measured in the nucleus, indicating a transition toward warmer gas. This scenario is supported by the results presented in paper~I, which showed that the photoelectric efficiency increases progressively with galactocentric distance, indicating more efficient gas heating toward larger radii. Although the overall radial variation remains modest, the change in the [C\textsc{ii}]$_{158}$/[O\textsc{i}]$_{63}$ ratio suggests subtle radial modifications in the physical conditions of the neutral ISM, possibly linked to variations in the hardness of the ionizing radiation field as traced by the increasing [Ne\textsc{iii}]$_{15}$/[Ne\textsc{ii}]$_{13}$ ratio.
The comparison with the photoionization models in Fig.~\ref{fig:CII_OI} and Fig.~\ref{fig:CII_appendix} shows that the observations are broadly consistent with ionization parameters in the range $-4 {\leq} \log(\rm U) {\leq} -3$ and hydrogen densities spanning $1 {\leq} \log(n_{\rm H}/cm^{-3}) {\leq} 2$, with NGC~6240 ranging from $2 {\leq} \log(n_{\rm H}/cm^{-3}) {\leq} 3$.
The inferred ages for the starburst component are compatible with 1--3.5\,Myr for solar metallicity, while super-solar metallicity models favor ages of $\sim$3.5\,Myr (see Fig.~\ref{fig:CII_OI} and Fig.~\ref{fig:CII_appendix}).  \\

\indent AGN-dominated galaxies.
The AGN-dominated nuclei are characterized by lower [C\textsc{ii}]$_{158}$/[O\textsc{i}]$_{63}$ ratios than starburst nuclei and extended regions, and they occupy a relatively narrow range centered around [C\textsc{ii}]$_{158}$/[O\textsc{i}]$_{63}{\approx}0.7$ that is largely independent of the inferred radiation field intensity (see the lower panels in Fig.~\ref{fig:CII_OI_radial_trend}). This behavior suggests that the neutral gas in AGN-dominated nuclei is typically warmer and denser than in starburst environments.
In contrast, in Fig.~\ref{fig:CII_OI}, the extended regions of moderate AGN hosts overlap with the parameter space occupied by starburst galaxies, and thus, they are characterized by a similar range of hydrogen densities. While this might suggest a transition toward lower-density neutral gas at larger galactocentric distances, a firm interpretation is limited by the small number of AGN-hosting sources and spatially resolved measurements.
NGC~7674, the most AGN-dominated system in our sample, is well reproduced by the AGN model grid for ionization parameters in the range $-3.5 {\leq} \rm log(U) {\leq} -3$ and hydrogen densities of $\rm log(n_{\rm H}/cm^{-3}) \mathord{\sim}3$ showing that for this galaxy, the neutral gas excitation might still be driven by the AGN feedback even to distances larger than 5~kpc from the nucleus (see Fig.~\ref{fig:CII_OI}). A similar trend was reported in paper~I: the intermediate regions exhibit PAH characteristics and radiation field conditions that are comparable to those of the nuclear regions, indicating that AGN activity can affect the gas and dust properties on kiloparsec scales, whereas the outermost regions display PAH properties that are characteristic of starburst systems, which is consistent with a declining AGN effect. However, in the \textit{Herschel grid} configuration, the statistics are very limited since we have data for only one extended region at large galactocentric distances. 

\section{Radial trends of the physical parameters}
\label{radial profile}

To interpret the physical conditions of the multiphase gas in the extreme environments of LIRGs, we relied on photoionization models to constrain the key parameters. In this section, we use measurements extracted with the \textit{Herschel grid}, which provide simultaneous access to MIR ([Ar\textsc{ii}]$_7$, [Ar\textsc{iii}]$_9$, [Ne\textsc{ii}]$_{13}$, [Ne\textsc{iii}]$_{15}$, [S\textsc{iii}]$_{19}$, and [S\textsc{iii}]$_{33}$) and FIR ([O\textsc{i}]$_{63}$, [O\textsc{iii}]$_{88}$, and [C\,\textsc{ii}]$_{158}$) diagnostics. We additionally made use of the \textit{Spitzer grid} configuration, which relies exclusively on MIR emission lines. For the two galaxies in our sample with the highest AGN fractions (NGC~5395 and NGC~7674), \textit{Spitzer} observations are available, while \textit{Herschel} data exist for NGC~7674 alone. The larger number of AGN-dominated data points in the \textit{Spitzer grid} reflects the broader spatial extent probed by the \textit{Spitzer} observations.\\
Although photoionization modeling inevitably relies on simplifying assumptions (related to geometry, dust chemical composition, or equilibrium conditions), these models still offer a reliable framework to explore the physical properties of gas exposed to an ionizing spectrum. By simultaneously comparing all available MIR and FIR line fluxes with predictions from a comprehensive grid of \texttt{CLOUDY} photoionization models, we can infer the underlying physical conditions that characterize the nuclei and extended regions of our LIRG sample. 
This comparison was carried out within a Bayesian statistical framework using the $\chi^2$ statistic as the optimization metric. For each combination of parameters in our model grid, we calculated

\begin{equation}
\chi^2 = \sum_{i=1}^{N} \frac{(\text{O}_i - \alpha \text{M}_i)^2}{\sigma_i^2},
\end{equation}
where $\rm O_i$ is the observed flux for emission line \textit{i}, 
$\rm M_i$ is the model-predicted flux, $\sigma_i$ is the observational uncertainty of the line flux, which includes an additional 15$\%$ to account for the sparse nature of the parameter grid, N is the number of available emission lines, and $\alpha$ is the scaling factor, which can be calculated analytically as

\begin{equation}
    \alpha = \frac{\sum_{i=1}^{N} \frac{O_i M_i}{\sigma_i^2}} {\sum_{i=1}^{N} \frac{M_i^2}{\sigma_i^2}}
\end{equation}
to avoid having to maximize the likelihood using normalized quantities, such as emission line ratios, and instead using absolute measurements of the line flux. 
The likelihood was then evaluated assuming Gaussian statistics, where the relative probability of each model is given by $\mathcal{P} = e^{(-\frac{\chi^2}{2})}$. We constructed one-dimensional posterior probability distributions for each parameter by marginalizing over all others, with the 50th percentile providing our best parameter estimate, and the 16th and 84th percentiles defining the 68$\%$ credible interval. To implement this approach, we developed and employed \texttt{SINEFIA}\footnote{Available on GitHub at \url{https://github.com/Sinefia/sinefia}}, a Python-based wrapper designed to generate \texttt{CLOUDY} model grids, and compare the grids with observational data. The tool automates the generation of model grids, runs \texttt{CLOUDY} within defined parameter spaces, and performs a Bayesian comparison with observations. It supports models of AGN and star-forming galaxies by allowing the ionizing source to be represented either by an AGN continuum or a stellar SED. This flexibility enables the consistent analysis of emission line spectra under different ionization conditions. For the starburst-dominated sample, we analyzed the radial distributions of the four best-fit parameters (ionization parameter, hydrogen density, stellar age, and metallicity) by grouping the individual nuclear and extended regions into bins of projected galactocentric distance. The uncertainty on each bin mean was calculated as $\sigma_{bin}=\sqrt{\sum_{i=1}^{\rm N} {\sigma_i}^2} /\rm N $, where $\sigma_{i}$ represents the individual uncertainties, and N is the number of measurements per bin. Because of their limited statistics, the AGN-hosting regions are shown individually and are not included in the linear fits.
The binned starburst measurements and the individual AGN measurements used in this analysis are provided in the electronic tables available at the CDS. 

\subsection{Variations in U as a function of galactocentric distance}
\label{U discussion}

\begin{figure*}[ht]
    \centering
    \begin{subfigure}[t]{0.49\textwidth}
        \centering
        \includegraphics[width=\linewidth]{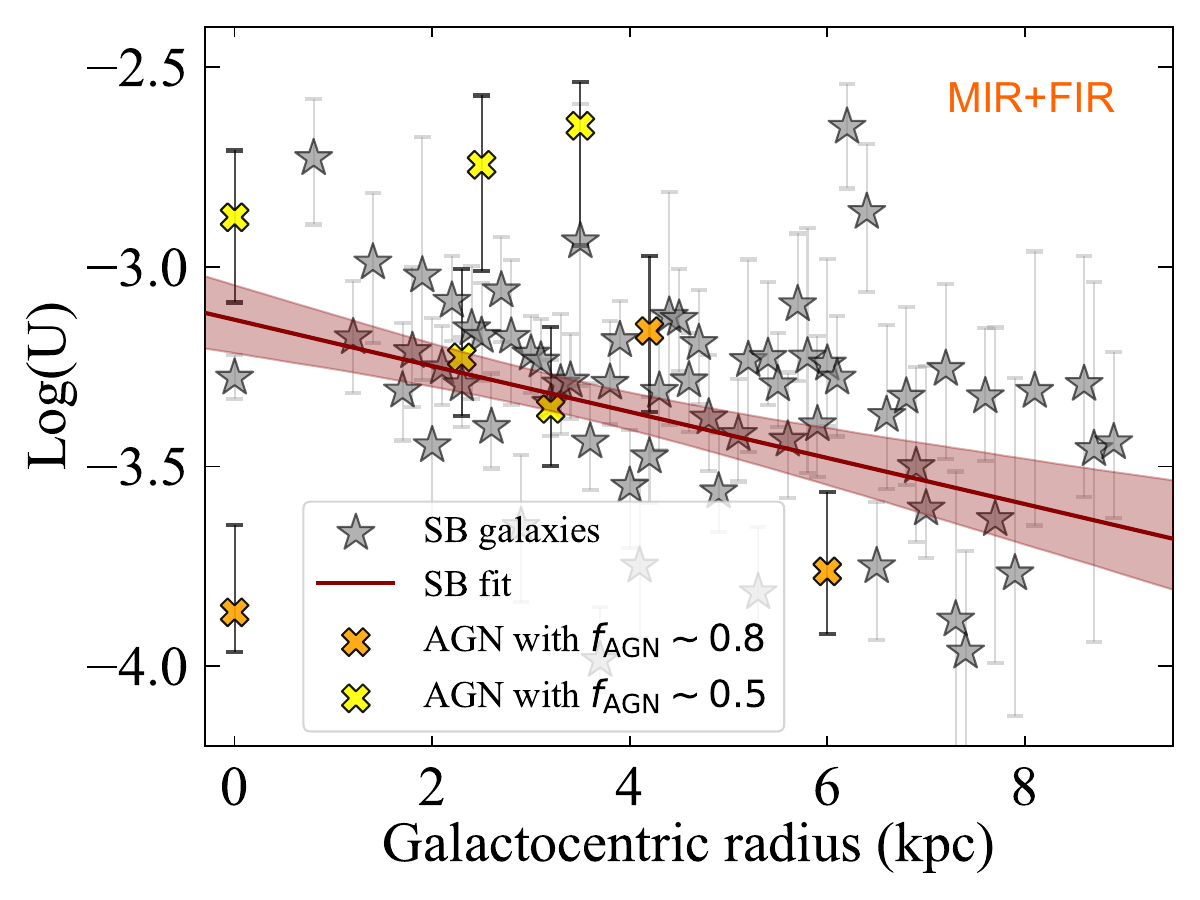}
        \vspace{-0.6 cm}
        
    \end{subfigure}
    \hfill
    \begin{subfigure}[t]{0.49\textwidth}
        \centering
        \includegraphics[width=\linewidth]{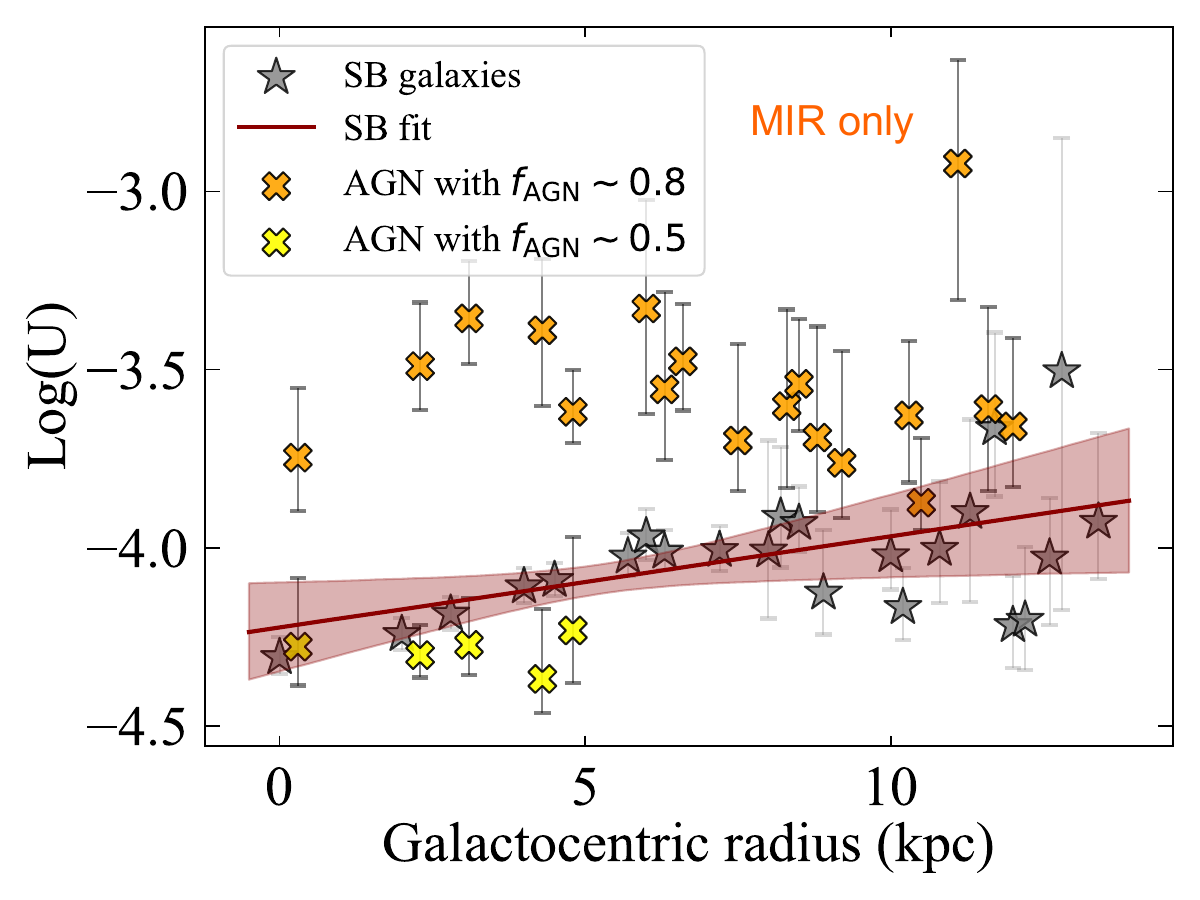}
        \vspace{-0.6 cm}
        
    \end{subfigure}
    \vspace{-0.3 cm}
    \caption{Radial distribution of the ionization parameter, $\log(U)$, as a function of projected galactocentric distance. The gray stars represent the mean values of the starburst-dominated regions within each galactocentric distance bin. Individual AGN-hosting regions with moderate AGN fractions ($f_{\rm AGN}{=}$0.5) are shown as yellow crosses, while those belonging to the most AGN-dominated galaxies ($f_{\rm AGN}{=}$0.8) are shown as orange crosses. The red linear relation in both panels is fitted exclusively to the binned starburst-dominated measurements. The AGN-hosting regions are shown only for comparison and are not included in either fit. \textit{Left:} Results derived using the \textit{Herschel grid} configuration, which combines MIR and FIR diagnostics. \textit{Right:} Results derived using the \textit{Spitzer grid} configuration, based exclusively on MIR diagnostics.}
    \label{fig:U_together}
\end{figure*}

Fig.~\ref{fig:U_together} presents the radial behavior of the ionization parameter, $\log(\rm U)$, as a function of projected galactocentric distance. The left panel shows measurements derived using the \textit{Herschel grid}, which combines MIR and FIR diagnostics, while the right panel displays results obtained with the \textit{Spitzer grid}, based exclusively on MIR emission lines. When the sample is separated according to the dominant power source, distinct radial behaviors emerge. \\

\indent Starburst galaxies.
When all available lines are used, the ionization parameter in the left panel of Fig.~\ref{fig:U_together} seems to decrease monotonically with radius. This is consistent with a centrally concentrated ionizing photon field and a progressive dilution of the radiation field toward larger galactocentric distances. To quantify this radial decline, we fit a linear relation to the starburst-dominated regions and estimated the uncertainties on the fit parameters using Monte Carlo sampling. The resulting best-fit relation is $\log(\rm U)  = -3.13^{+0.04}_{-0.04} -0.058^{+0.01}_{-0.01} \times r_{gal}$, where $r_{gal}$ is the galactocentric distance.
Several integral-field spectroscopic studies of LIRGs have revealed pronounced spatial variations in gas excitation within merging systems, with nuclear regions typically dominated by star formation or AGN photoionization, and extranuclear regions exhibiting different excitation regimes, often affected by shocks (e.g., \citet{Monreal10, West12, Colina15}). Complementary constraints from galaxy-integrated diagnostics further indicate that LIRGs hosting compact nuclear star formation exhibit systematically higher ionization parameters than more extended systems (\citet{Gracia11}). This behavior is commonly interpreted in terms of dust-bounded H\textsc{ii} regions and intense central radiation fields. Our spatially resolved measurements provide direct observational support for this interpretation by explicitly quantifying how the ionization parameter varies with galactocentric radius in starburst-dominated LIRGs. 

A markedly different behavior emerges when the ionization parameter is derived using only MIR diagnostics, as shown in the right panel of Fig.~\ref{fig:U_together}. In contrast to the declining radial trend observed when combining MIR+FIR lines with the \textit{Herschel grid} for starburst galaxies, the \textit{Spitzer grid} measurements exhibit an overall increase in $\log(\rm U)$ with galactocentric radius. Such a trend is counterintuitive for starburst galaxies, as it would imply systematically harder or more intense radiation fields at large distances from the nucleus than in the central regions. We further discuss this apparent contradiction at the end of this subsection. \\

\indent AGN-dominated galaxies.
The AGN-hosting systems seem to show different qualitative behaviors. In the left panel of Fig.~\ref{fig:U_together}, galaxies with moderate AGN fractions exhibit an approximately constant radial profile of log(U) over the full radial extent we probed. In contrast, galaxies dominated by powerful AGNs show a non-monotonic radial trend.
Previous spatially resolved studies of AGN-hosting galaxies indicated that the radial trends of the ionization parameter are not universal. Radial profiles derived for narrow-line regions of nearby Seyfert galaxies show that the ionization parameter does not necessarily peak at the nucleus. In some cases, the maximum value is reached at off-nuclear locations, followed by a decline at larger radii \citep{Bennert06}. These departures from a simple monotonic decrease are attributed to effects such as nuclear obscuration and the geometry of the ionization cones. Other studies of extended narrow-line regions reported that ionization conditions can remain approximately constant over several kiloparsecs within AGN ionization cones but can decrease with distance in regions outside the cones \citep{Cracco11}.

\begin{figure*}[t]
    \centering

    \begin{minipage}[t]{0.32\textwidth}
        \centering
        \textbf{(a)}\\
        \includegraphics[width=\linewidth]
        {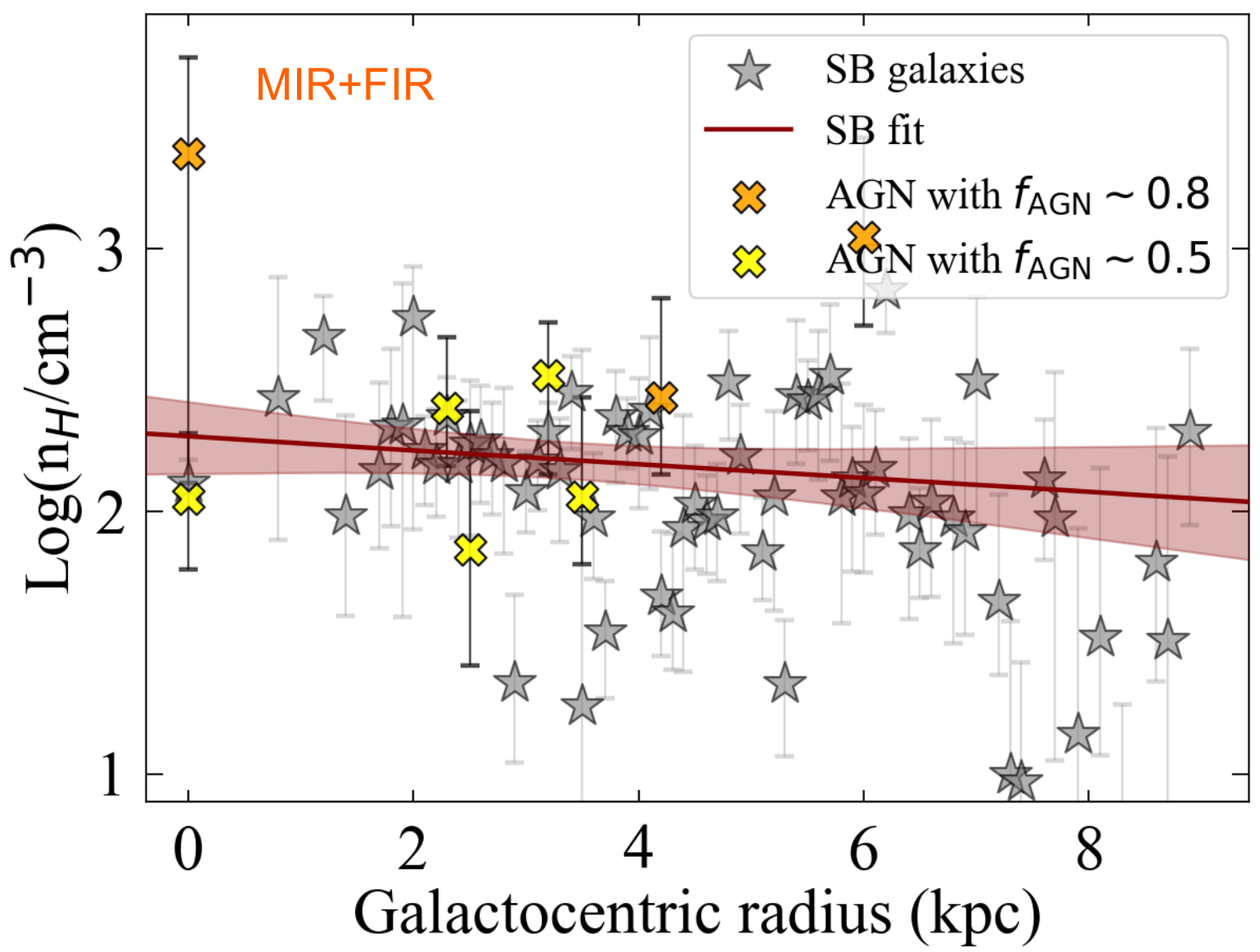}
    \end{minipage}
    \hfill
    \begin{minipage}[t]{0.32\textwidth}
        \centering
        \textbf{(b)}\\
        \includegraphics[width=\linewidth]
        {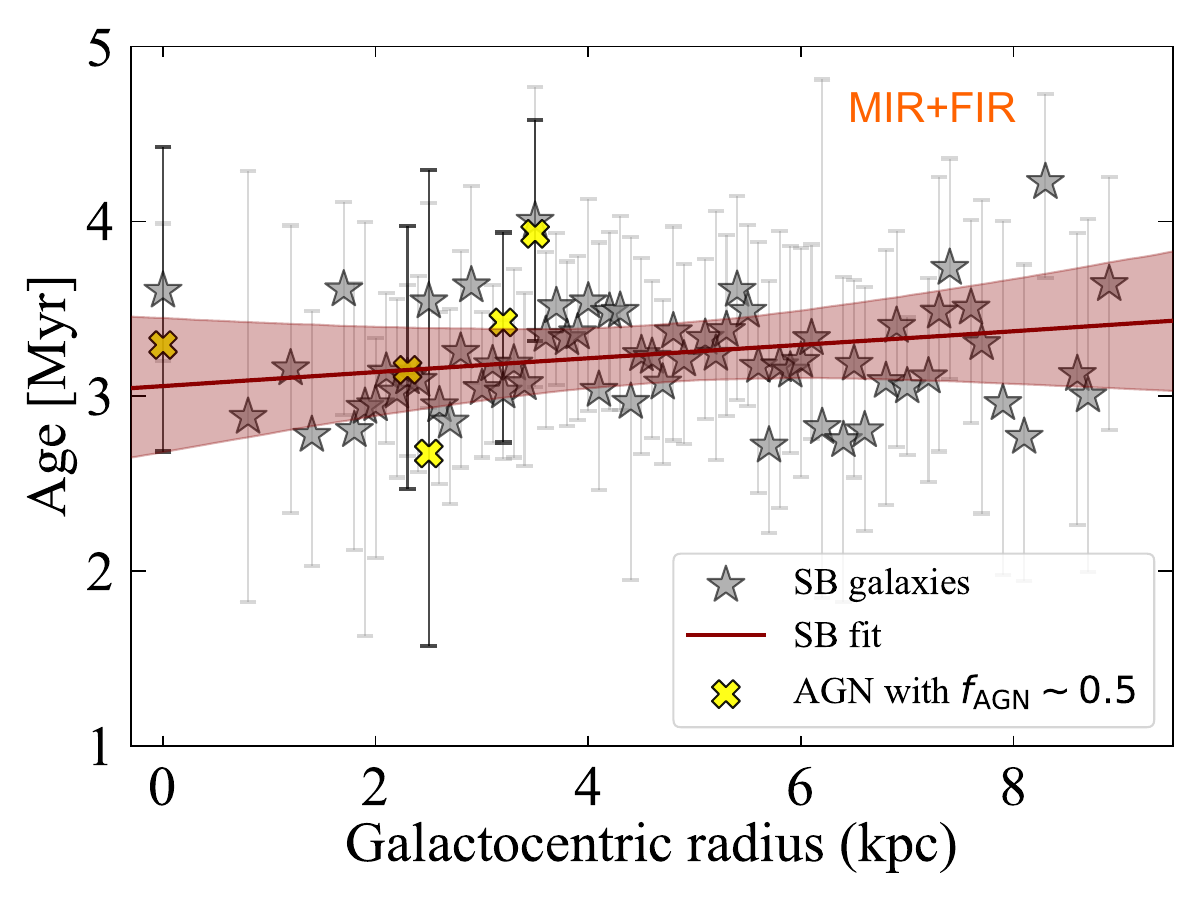}
    \end{minipage}
    \hfill
    \begin{minipage}[t]{0.32\textwidth}
        \centering
        \textbf{(c)}\\
        \includegraphics[width=\linewidth]
        {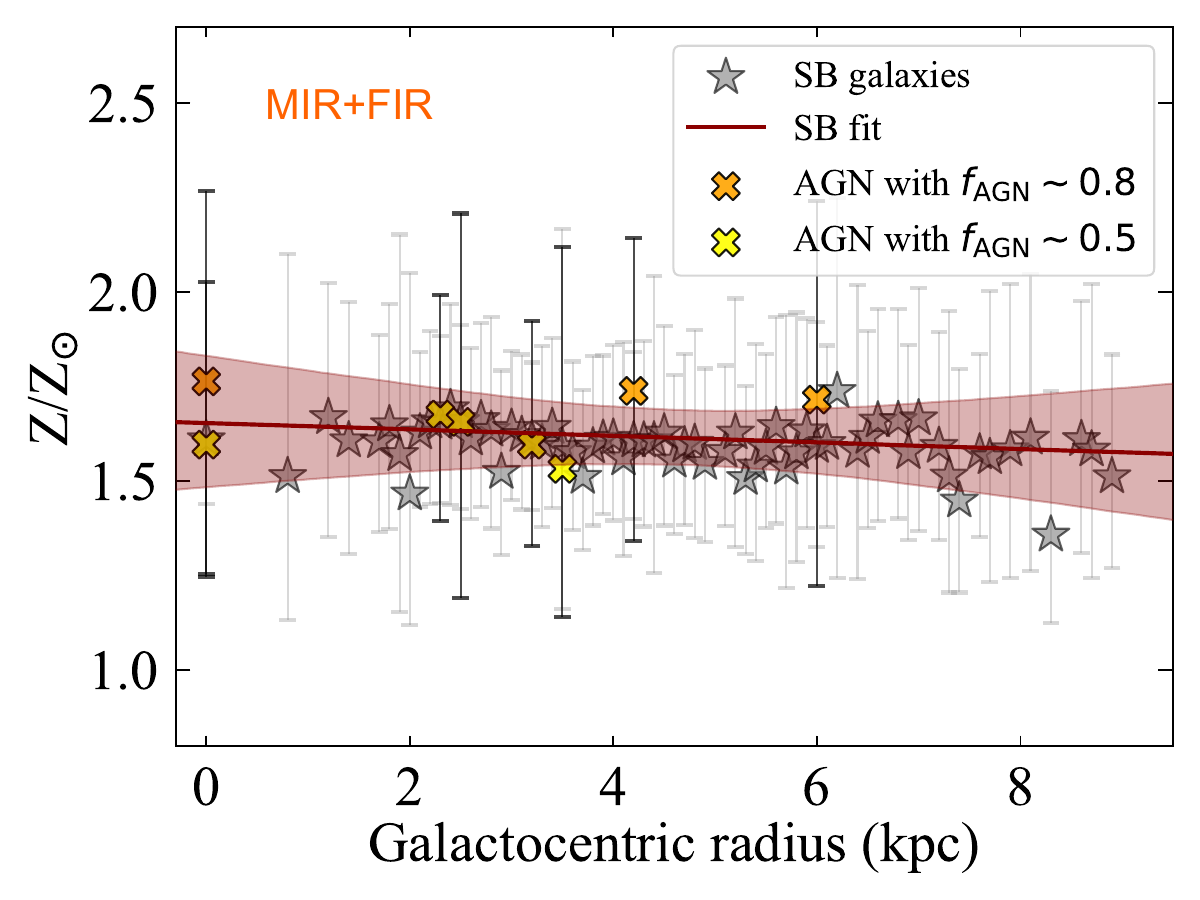}
    \end{minipage}

    \caption{Radial distributions of the physical parameters as functions
    of projected galactocentric distance for the \textit{Herschel grid}
    configuration. Panel a: Radial profile of the hydrogen density,
    $\log(n_{\rm H})$. Panel b: Radial profile of the stellar age.
    Panel c: Radial profile of the gas metallicity.}
    \label{fig:physical_parameters_Herschel}
\end{figure*}

The range of trends observed in our AGN subsample, including the nearly flat radial profiles at moderate AGN fractions and the non-monotonic behavior in AGN-dominated sources, thus reflects the complex interplay between nuclear activity and spatially extended ionized gas in LIRGs. Larger samples of spatially resolved fine-structure line measurements in AGN-dominated galaxies are needed to establish statistically robust constraints on how the intensity and hardness of the radiation field vary with distance from the central power source. 

The discrepancy between MIR-inferred and MIR- plus FIR-inferred ionization parameters suggests that ionization parameter values inferred solely from MIR line ratios do not fully capture the radial ionization structure of LIRGs. Additional tests showed that the inversion of the radial trend between the two datasets (MIR plus FIR lines, and MIR lines alone) is primarily driven by the inclusion of the [C\textsc{ii}]$_{158, \rm PDR}$ line in the \textit{Herschel grid} configuration. 
Since [CII]$_{158}$ emission arises from the ionized and neutral phases of the ISM, the [C\textsc{ii}]$_{158, \rm PDR}$ was computed by following the prescriptions of \citet{Tanio17} and \citet{McKinney21}, and we calculated the PDR-associated [CII] luminosity according to

\begin{equation}
L_{\rm [C\,II],PDR}
=
\rm f_{\rm PDR} \times L_{\rm [C\,II],total},
\end{equation}

where $f_{\rm PDR}$ denotes the fraction of the total [C~II] luminosity produced by neutral PDR gas. We derived $f_{\rm PDR}$ from the FIR continuum color $S_{63}/S_{158}$ measured on the Herschel grid, adopting the empirical relation presented by \citet{Tanio17}.
The [C\textsc{ii}]$_{158,\rm PDR}$ emission, which dominates the cooling of neutral gas in PDRs, strongly constrains the ionization parameter in the nuclear regions, favoring higher U values toward the center and lower values at large radii. When this constraint is removed, as in the MIR-only analysis, the resulting solutions allow systematically lower U in the nucleus and higher U at larger distances, reflecting the increasing  radial trends observed in Fig.~\ref{fig:Ne_SIII_radial_trend}.

These results demonstrate that FIR diagnostics, and in particular [C\textsc{ii}], play a critical role in anchoring the ionization conditions of the multiphase ISM in LIRGs. Analyses based exclusively on MIR emission lines might thus lead to misleading radial trends in the ionization parameter, especially in spatially resolved studies. The combined use of MIR and FIR tracers is essential to recover physically meaningful radial profiles of U and to properly account for the coupling between ionized and neutral gas across galactic disks.

\subsection{Variations in n$_H$ as a function of galactocentric distance}

Panel a of Fig.~\ref{fig:physical_parameters_Herschel} presents the radial distribution of the average hydrogen density, $\log(\rm n_{\rm H})$, as a function of projected distance from the galaxy center. As for the ionization parameter, we fit a linear function to the starburst-dominated regions. The panel shows measurements derived using the \textit{Herschel grid}, while the panel displaying results obtained with the \textit{Spitzer grid}, and thus based exclusively on MIR emission lines, is available in Sect.~\ref{Spitzer models}. \\

\indent Starburst galaxies.
The total hydrogen density remains approximately constant with radius, with typical values of $\log(\rm n_{\rm H}/\mathrm{cm^{-3}}){\sim}2.3$. Previous FIR spectroscopic studies have provided important constraints on the physical conditions of PDRs in star-forming galaxies and LIRGs. Using galaxy-integrated measurements of the main PDR cooling lines, such as [C\textsc{ii}] and [O\textsc{i}], \citet{Malhotra01} showed that the FIR line ratios correlate with the dust temperature and can be interpreted with PDR models to infer the gas densities ($10^2~ \rm cm^{-3} {\leq} n {\leq} 10^{4.5}~\rm cm^{-3}$) and FUV radiation fields ($10^2 {\leq} G_0 {\leq} 10^{4.5}$) in nearby star-forming galaxies. Herschel/PACS observations made it possible to extend these analyses to LIRGs. \citet{Tanio14} showed that [C\textsc{ii}]$_{158}$ emission in local LIRGs is spatially extended over kiloparsec scales, indicating that the FIR line emission arises not only from their nuclei, but also from extended PDRs. Building on this work, \citet{Tanio17} used a homogeneous Herschel/PACS dataset to derive average PDR hydrogen densities and radiation field intensities for individual GOALS galaxies by comparing the observed FIR line ratios with PDR models. Their analysis, which explicitly focused on systems dominated by star formation, found that typical PDR hydrogen densities in LIRGs have an average of $\mathord{\sim}500~\mathrm{cm^{-3}}$. This is consistent with the characteristic densities measured for the starburst galaxies in our sample. \\

\indent AGN-dominated galaxies.
For the AGN-hosting systems, the interpretation of radial variations in the hydrogen density is more uncertain than for the starburst-dominated population, primarily because the spatially resolved measurements, particularly in the combined MIR+FIR grid, are affected by larger uncertainties. Galaxies with moderate AGN fractions show values of $\log(n_{\rm H}/{\rm cm^{-3}}){\sim}2.3$, which are similar to those measured in starburst-dominated systems. In contrast, the trend in the most AGN-dominated source is non-flat. However, we emphasize that these radial variations are not statistically significant because all measurements remain consistent within the $1\sigma$ uncertainties. Consequently, the apparent non-monotonic behavior should be interpreted with caution. Our measurements therefore provide tentative insight into how the PDR hydrogen density behaves in AGN-hosting LIRGs and how it differs from the starburst-dominated population. Larger samples of spatially resolved observations are required to confirm these trends, however.

For starburst- and AGN-dominated systems, the hydrogen-density profiles derived using the \textit{Spitzer grid} (see Fig.~\ref{fig:physical_parameters_Spitzer}, panel a) are broadly consistent with those obtained from the \textit{Herschel grid}. Although the \textit{Spitzer}-based measurements rely exclusively on MIR diagnostics, the resulting $\log(n_{\rm H})$ values fall within or very close to the uncertainties associated with the \textit{Herschel}-based fits.

\subsection{Variations in the stellar age as a function of galactocentric distance}

\label{age_profile}

Panel b of Fig.~\ref{fig:physical_parameters_Herschel} shows the radial distribution of the stellar ages as a function of projected distance from the galaxy center as derived using the \textit{Herschel grid}. Although significant scatter is present, extended regions of starburst galaxies tend to cluster around ${\sim}$3~Myr, regardless of the galactocentric distance. Previous studies have shown that intense star formation in LIRGs occurs on short timescales and is therefore dominated by young stellar populations. Using spatially resolved near-IR spectroscopy of nearby starburst galaxies, \citet{Forster04} derived characteristic starburst ages of a few million years and found no strong evidence of age gradients within the starburst regions. This agrees with our findings. Similarly, \citet{Inami13} showed that the MIR fine-structure line ratios observed in GOALS starburst-dominated galaxies are best reproduced by photoionization models assuming young stellar populations with characteristic ages of a few million years (typically ${\sim}3$--4~Myr). Together, these studies support a picture in which the IR output of starburst-dominated LIRGs is driven by recent star formation that extends throughout the entire extent of the galaxies. This agrees with the uniformly young ages inferred in the regions probed in our analysis. 

For starburst- and AGN-hosting systems, the stellar age profiles derived using the \textit{Spitzer grid} (see Fig.~\ref{fig:physical_parameters_Spitzer}, panel b) are consistent with those obtained from the \textit{Herschel grid}. The agreement between the two independent configurations indicates that the inferred stellar ages are robust against the choice of IR diagnostics and spatial sampling. This reinforces the conclusion that the regions we probed are dominated by uniformly young stellar populations.

\subsection{Variations in metallicity as a function of galactocentric distance}
\label{Z_profile}

Panel c of Figure~\ref{fig:physical_parameters_Herschel} shows the radial behavior of the gas-phase metallicity as a function of projected galactocentric distance as derived using the \textit{Herschel grid}. \\

\indent Starburst galaxies.
Starburst-dominated galaxies exhibit flat radial metallicity profiles, with values clustering around $Z/Z_\odot\mathord{\sim}1.6$. This behavior is consistent with previous studies, which showed that galaxy interactions and mergers efficiently flatten preexisting abundance gradients. Using optical emission line diagnostics, \citet{Rupke10} demonstrated that interacting galaxies exhibit significantly shallower gas-phase metallicity gradients than isolated disk galaxies. This effect was attributed to radial gas flows and mixing driven by tidal interactions. Similar conclusions were reached by \citet{Kewley10}, who showed that mergers can rapidly erase metallicity gradients even at early interaction stages. Integral-field observations of GOALS LIRGs by \citet{Rich12} further demonstrated that the metallicity gradients become progressively flatter with merger stage, reflecting large-scale gas mixing during interactions. These results provide a natural explanation for the absence of strong radial metallicity variations in merging starburst-dominated systems. \\

\indent AGN-dominated galaxies.
The AGN-hosting galaxies in our sample exhibit approximately flat metallicity profiles. The strongly AGN-dominated system shows a slightly higher metallicity than the starburst population, with $Z/Z_\odot\mathord{\sim}1.75$, whereas the metallicity values of the moderately AGN-dominated galaxy are consistent with those of starburst systems. Observational studies of active galaxies indicate that the nuclear metallicities can differ from those inferred for purely star-forming regions. Using spatially resolved SDSS–MaNGA data, \citet{Nascimento22} found that the metallicity of the narrow-line regions in Seyfert galaxies is often offset from the extrapolated central metallicity of the host galaxy disk, suggesting that gas inflows or nuclear processes can modify the chemical properties of the circumnuclear gas. 

For starburst- and AGN-dominated galaxies, the metallicity profiles derived from the \textit{Spitzer grid} (see Fig.~\ref{fig:physical_parameters_Spitzer}, panel c) are fully consistent with those obtained using the \textit{Herschel grid}. This agreement indicates that the inferred metallicities and their lack of radial variation are robust against the choice of IR diagnostics and spatial sampling.

\section{Summary and conclusions}
\label{conclusions}

We have presented a spatially resolved multiphase study of the ISM in a representative sample of 67 nearby LIRGs from GOALS, for which we combined MIR spectroscopy from Spitzer/IRS with FIR line maps from Herschel/PACS. By extracting spectra in square-aperture grids out to projected galactocentric distances of $\sim$14\,kpc, we quantified how the ionized- and neutral-gas diagnostics and the physical parameters inferred from photoionization modeling vary with radius in both starburst- and AGN-dominated systems. The main results are listed below.

\begin{enumerate}
    \item Starburst regions show an overall increase in the [S\textsc{iii}]$_{33}$/[S\textsc{iii}]$_{19}$ ratio from $\mathord{\sim}1$ to $\mathord{\sim}2$ out to $\mathord{\sim}$8\,kpc. This is consistent with a radially decreasing electron density. At larger radii, the ratio displays increased scatter and not a monotonic trend, which might reflect a broader range of radiation field conditions in the outer regions, as [Ne\textsc{iii}]$_{15}$/[Ne\textsc{ii}]$_{13}$ increases with galactocentric distance (from ${\sim}$ 0.2 to ${\sim}$ 0.6). In the [S\textsc{iii}]$_{33/19}$ versus [Ne\textsc{iii}]$_{15}$/[Ne\textsc{ii}]$_{13}$ plane, the hydrogen density spans of $1 {\lesssim} \log (n_{\rm H}/cm^{-3}) {\lesssim} 3$ and a broad range of radiation hardness (for the nuclear regions $\rm \log(U){\sim}{-}4$ to $\rm \log(U){\sim}{-}3.5$, and for the extended regions $\rm \log(U){\sim}{-}4$ to $\rm \log(U){\sim}{-}3$) at 3.5~Myr and Z$_{\odot}$. 
    The AGN-dominated nuclei extend toward harder and more diverse excitation conditions than the starburst-dominated regions, with NGC~7674 reaching the highest [Ne\textsc{iii}]$_{15}$/[Ne\textsc{ii}]$_{13}$ and [S\textsc{iii}]$_{33}$/[S\textsc{iii}]$_{19}$ ratios in the sample. The comparison with the CLOUDY grids suggests ionization parameters in the range $\log(U)\sim-4$ to $-3$ and densities of $2\lesssim\log(n_{\rm H}/{\rm cm^{-3}})\lesssim3$ for most AGN nuclei, although the NGC~7674 nucleus is not fully reproduced by the models. The extended regions of most AGN hosts move closer to the starburst locus, suggesting a reduced AGN effect at larger galactocentric distances. However, NGC~7674 remains distinct out to $\sim$5~kpc, showing nuclear-like [Ne\textsc{iii}]$_{15}$/[Ne\textsc{ii}]$_{13}$ ratios and lower [S\textsc{iii}]$_{33}$/[S\textsc{iii}]$_{19}$ values that are consistent with harder radiation fields and denser ionized gas. The median radial trends further suggest that the strongest AGNs maintain a higher excitation than the moderate-AGN systems over most of the radial range, although this comparison should be interpreted cautiously given the limited AGN statistics. \\
    
    \item The [C\textsc{ii}]$_{158}$/[O\textsc{i}]$_{63}$ ratio exhibits only modest radial variations in starburst galaxies. Although these trends are not statistically significant, they suggest that the physical conditions of the neutral ISM might be related to changes in the hardness of the radiation field with galactocentric distance. The comparison with photoionization models indicates ionization parameters in the range $-4 {\lesssim} \log U {\lesssim} -3$ and hydrogen densities spanning $1 {\lesssim} \log (n_{\rm H}/cm^{-3}) {\lesssim} 2$. The AGN-dominated nuclei occupy a narrow locus at systematically lower [C\textsc{ii}]/[O\textsc{i}] values ($\mathord{\sim}0.6$), which is consistent with warmer and denser neutral gas. Their extended regions generally overlap the starburst regime, indicating a declining AGN effect with increasing galactocentric distance. NGC~7674 remains an exception and maintains AGN-like excitation out to $\sim$5\,kpc, but this result is based on a very limited number of data points. It is well reproduced by the AGN model grid for ionization parameters in the range $-3.5 {\leq} \rm log(U) {\leq} -3$ and hydrogen densities of $\rm log(n_{\rm H}/cm^{-3}) \mathord{\sim}3$, showing that in the strongest AGN of our sample, the neutral gas excitation might still be driven by the AGN feedback even to distances larger than 5~kpc from the nucleus. \\ 
    
    \item The ionization parameter profiles derived from MIR$+$FIR diagnostics (the \textit{Herschel grid}) show a monotonic decline of the ionization parameter ($\log(\rm U)  = -3.13^{+0.04}_{-0.04} -0.058^{+0.01}_{-0.01} \times r_{gal}$) in starbursts, approximately flat profiles in moderate-AGN systems ($\rm log(U) {\approx}~{-}2.75$), and a non-monotonic behavior in strongly AGN-dominated sources. In contrast, MIR-only modeling (the \textit{Spitzer grid}) yielded an inverted trend for starbursts. This inversion is driven primarily by the absence of [C\textsc{ii}]$_{158, \rm PDR}$, demonstrating that FIR constraints are essential to anchor the ionization conditions of the multiphase ISM and to avoid misleading radial U trends in spatially resolved studies. \\ 
    
    \item Starburst galaxies display approximately constant $\log(\rm n_{\rm H})$ with radius (typical $\log(n_{\rm H}/{\rm cm^{-3}})\mathord{\sim}2.3$). In AGN-hosting systems, two distinct behaviors are observed. Galaxies with moderate AGN fractions show only weak radial variations in $n_{\rm H}$, with values generally comparable to those of starburst galaxies. In contrast, the strongly AGN-dominated source exhibits enhanced central densities and larger scatter toward larger radii. Although these trends are not statistically significant owing to the limited number of spatially resolved measurements, they suggest that additional physical processes, beyond pure photoionization, might affect the neutral gas in AGN environments.
    Unlike $\log(\rm U)$, the inferred $\log(\rm n_{\rm H})$ profiles are broadly consistent between MIR-only and MIR plus FIR analyses. \\
    
    \item The inferred ages for the starburst component are uniformly young (typically a few million years, clustering around $\sim$3\,Myr) and do not significantly depend on the radius. The consistency between the \textit{Herschel} and \textit{Spitzer} grid configurations indicates that these age constraints are robust against the choice of IR diagnostics. \\
    
    \item The gas-phase metallicities are approximately flat with radius over kiloparsec scales, with starburst and moderate AGN-dominated galaxies clustering around $Z/Z_\odot\mathord{\sim}$1.6, whereas the strong AGN host shows a slightly higher normalization ($Z/Z_\odot\mathord{\sim}$1.75). The metallicities derived from MIR-only and MIR plus FIR fits agree, indicating that the lack of strong gradients is robust to the diagnostic set.
\end{enumerate}

Overall, this work demonstrates that the properties of multiple ISM phases in luminous dusty galaxies evolve systematically with galactocentric distance, reflecting the changing effect of star formation and AGN activity. They also show that it is critical to combine MIR and FIR diagnostics for accurately recovering physically meaningful ionization parameter trends. These results provide a benchmark for future high-resolution studies with the JWST Mid-Infrared Instrument (MIRI) and the Atacama Large Millimeter/submillimeter Array (ALMA), which will enable the same multiphase approach to be extended to smaller spatial scales and to disentangle the relative roles of AGN radiation, star formation, and mechanically driven processes in shaping the ISM of dusty galaxies.

\section{Data availability}
The tabulated measurements underlying the radial profiles, diagnostic diagrams, and photoionization-model results are available in electronic form at the CDS via anonymous ftp to \texttt{cdsarc.cds.uniTablesstra.fr} (130.79.128.5) or via
\url{https://cdsarc.cds.unistra.fr/cgi-bin/qcat?J/A+A/}.

\begin{acknowledgements}

\\ CML and MSG acknowledge the research project was supported by the Hellenic Foundation for Research and Innovation (HFRI) under the "2nd Call for HFRI Research Projects to support Faculty Members $\&$ Researchers" (Project Number: 03382).
CML wishes to thank the "Summer School for Astrostatistics in Crete" for providing training on the statistical methods adopted in this work. 
MB acknowledges support from the Juan de La Cierva scholarship with reference JDC2023- 052684 -I, funded by MICIU/AEI/10.13039/501100011033 y por el FSE+ and from the Agencia Estatal de Investigacion of the Ministerio de Ciencia, Innovacion y Universidades (MCIU/AEI) under the grant ``Tracking active galactic nuclei feedback from parsec to kiloparsec scales'', with reference PID2022 $-$ 141105NB $-$ I00 and the European Regional Development Fund (ERDF).
DBS gratefully acknowledges support from NSF Grant 2407752.
The National Radio Astronomy Observatory and Green Bank Observatory are facilities of the U.S. National Science Foundation operated under cooperative agreement by Associated Universities, Inc.
IS acknowledges fundings from the European Research Council (ERC) DistantDust (Grant No.101117541) and the Atracc\'{i}on de Talento Grant No.2022-T1/TIC-20472 of the Comunidad de Madrid, Spain.
CR acknowledges support from SNSF Consolidator grant F01$-$13252 and ANID BASAL project FB210003.

\end{acknowledgements}

\bibliographystyle{aa}

\bibliography{biblio}

\clearpage

\begin{appendix}
\onecolumn
\raggedbottom

\setlength{\intextsep}{8pt plus 2pt minus 2pt}
\setlength{\floatsep}{8pt plus 2pt minus 2pt}
\setlength{\textfloatsep}{8pt plus 2pt minus 2pt}
\captionsetup{skip=4pt}


\section{CAFE fits}
\label{cafe_fits}

\begin{figure}[H]
    \centering
    \includegraphics[width=0.44\textwidth]{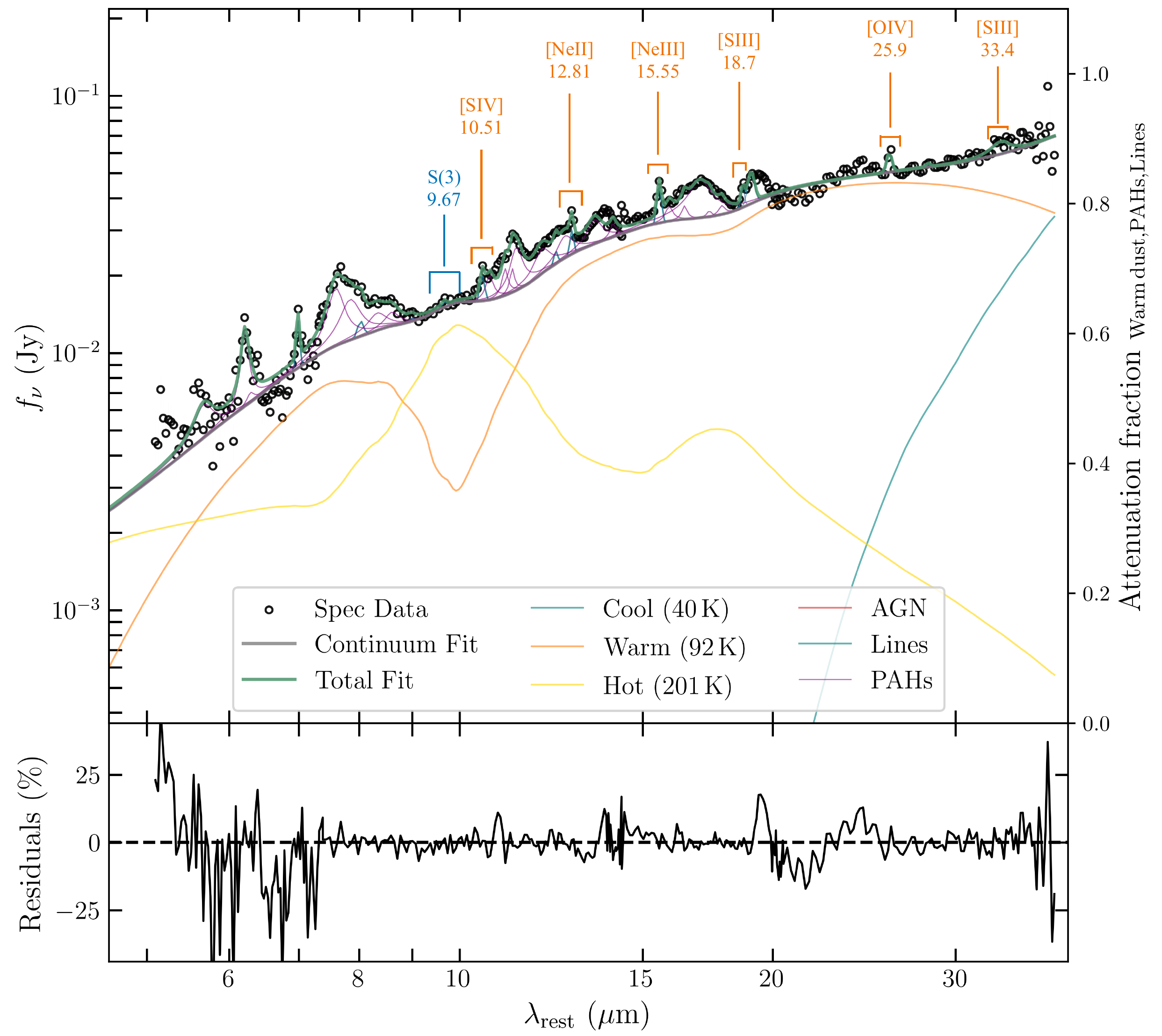}
    \hfill
    \includegraphics[width=0.44\textwidth]{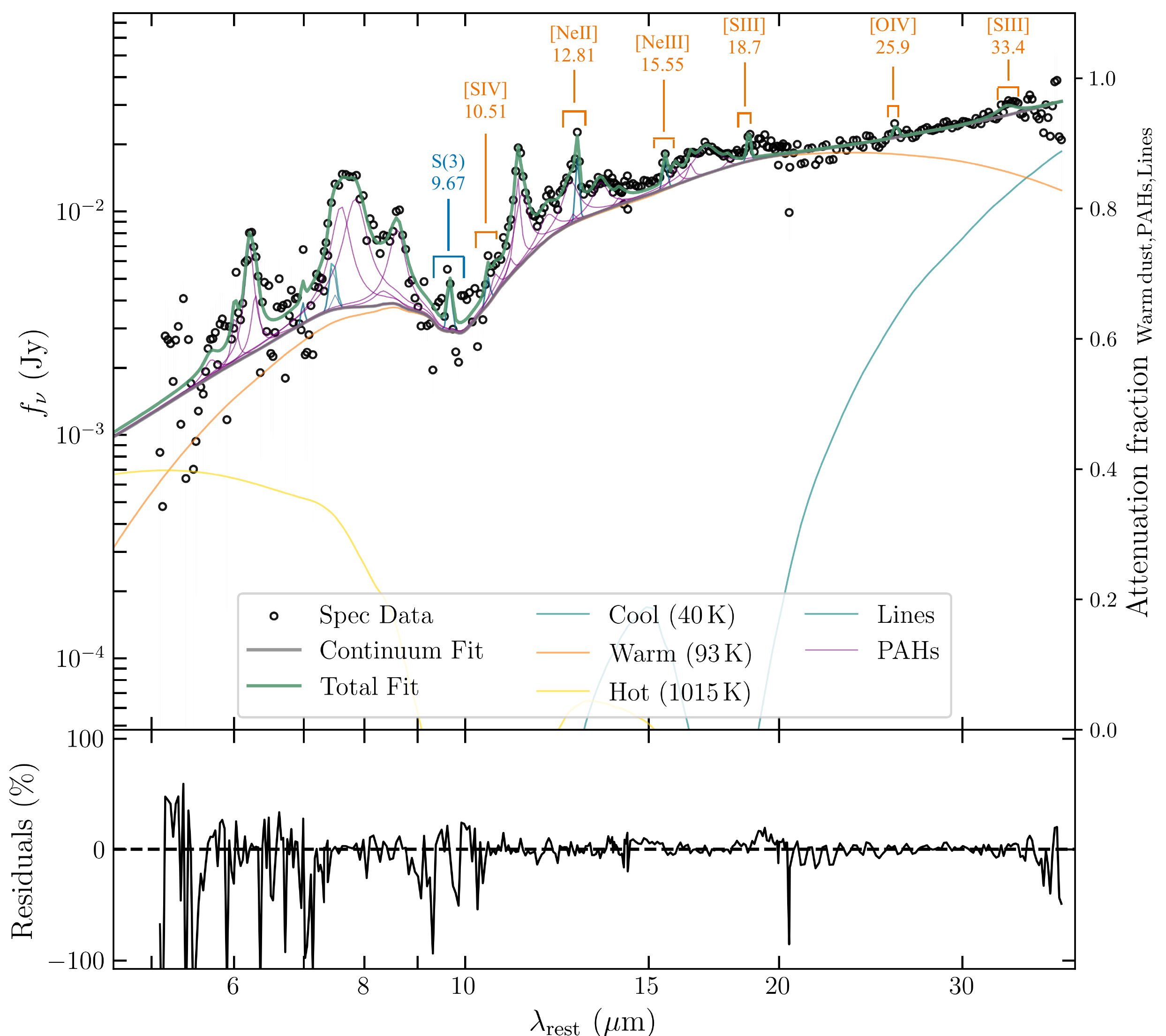}

    \caption{MIR spectra extracted from two spatially distinct regions
    of NGC~7674. The left panel shows the nuclear region, while the
    right panel shows one of the most distant regions from the nucleus
    included in this study. The observed spectra are shown as black
    points, while the green curves represent the total model fits.
    The gray curves trace the dust continuum, composed of cool (cyan),
    warm (orange), and hot (yellow) components. The PAH features are
    shown in purple, and the atomic and molecular emission lines are
    indicated in blue.}
    \label{fig:cafe_fits}
\end{figure}

As a representative illustration of the spectral decomposition,
Fig.~\ref{fig:cafe_fits} shows MIR spectra of NGC~7674 extracted
from two spatially distinct regions within the galaxy. The left
panel corresponds to the nuclear region, while the right panel
presents the spectrum of one of the most distant apertures included
in our analysis. The comparison between these two regions
highlights the spectral differences associated with their location
within the galaxy and demonstrates the ability of the CAFE fitting
framework to reproduce the observed MIR emission across a wide
range of physical environments.

\Needspace{0.32\textheight}
\section{PopStar SED}
\label{popstar}

\begin{figure}[H]
    \centering
    \includegraphics[width=0.55\textwidth]{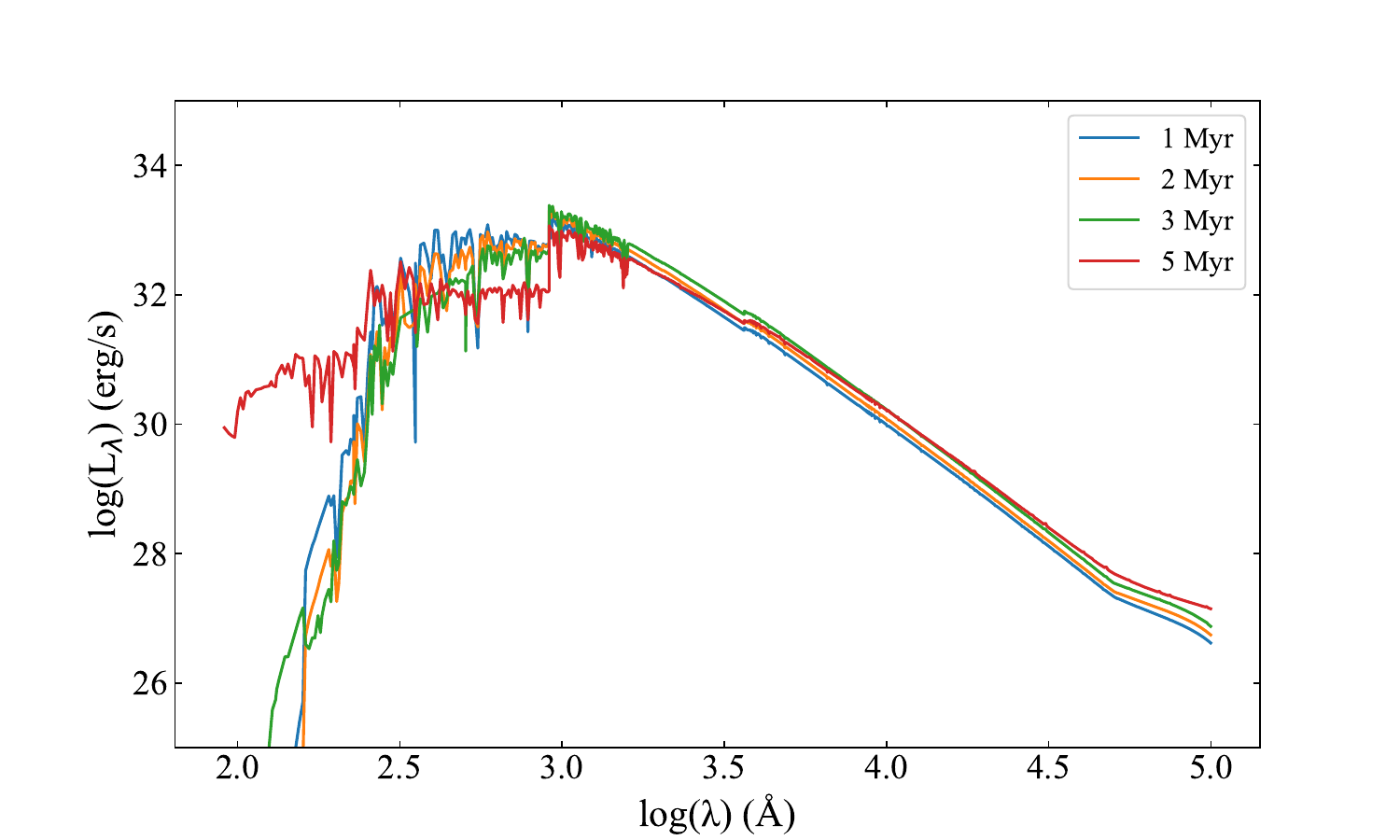}

    \caption{SEDs from PopStar models showing the evolution of
    stellar populations at different ages: 1, 2, 3, and 5 Myr.}
    \label{fig:Popstar}
\end{figure}


\Needspace{0.42\textheight}
\section{Additional line ratios and photoionization model plots}
\label{Additional models Appendix}

In the main text, we show the CLOUDY photoionization model grids
computed for a stellar age of 3.5 Myr at both metallicities,
$Z=Z_{\odot}$ and $Z=2.5Z_{\odot}$. In this Appendix, we complement
those diagnostics by presenting the corresponding grids for younger
and older stellar populations, with ages of 1 and 5 Myr. Symbols
follow the same convention as in Fig.~\ref{fig:SIII}.

\begin{figure}[H]
    \centering
    \includegraphics[width=0.39\textwidth]
    {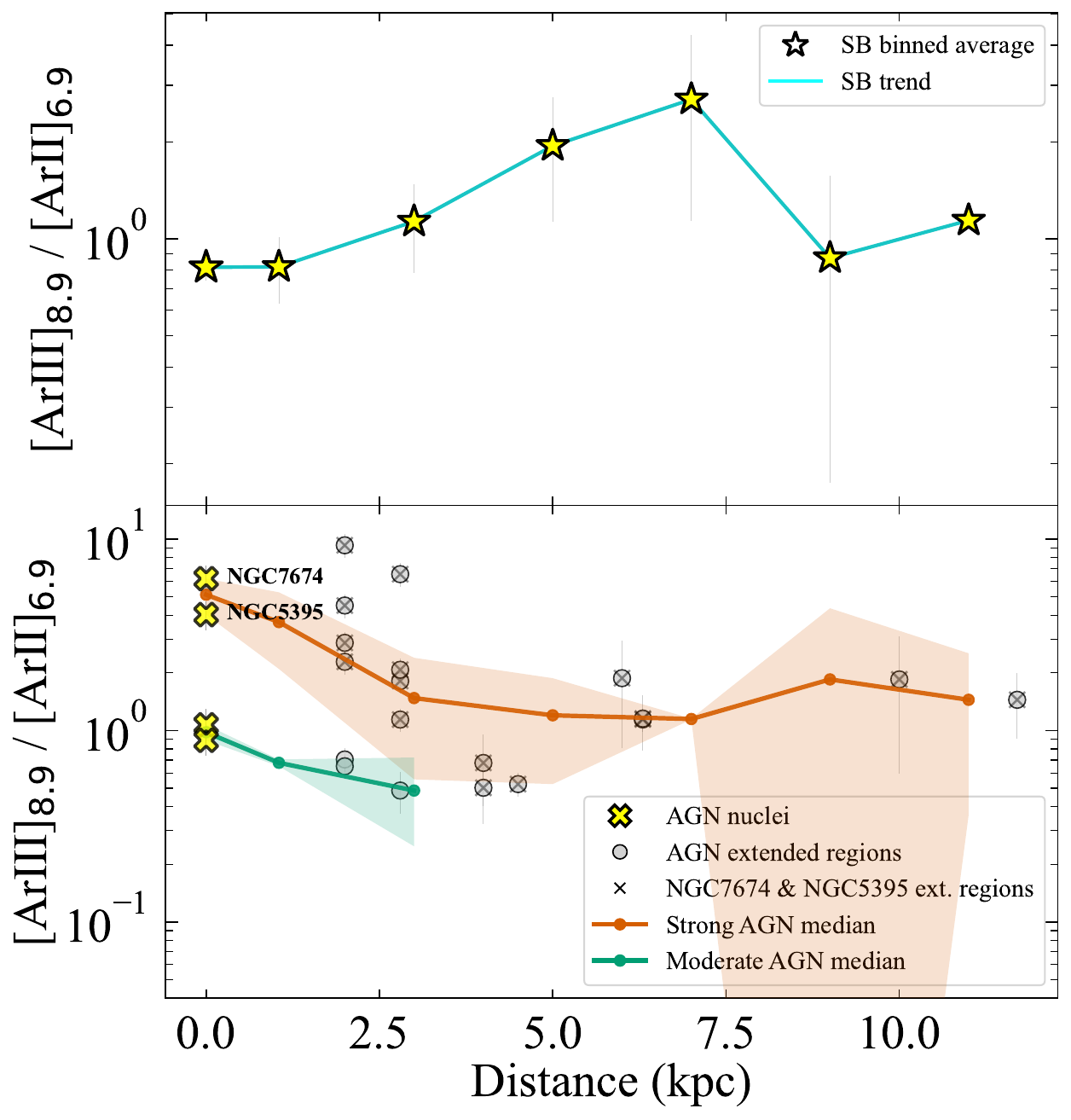}
    \hspace{0.025\textwidth}
    \includegraphics[width=0.39\textwidth]
    {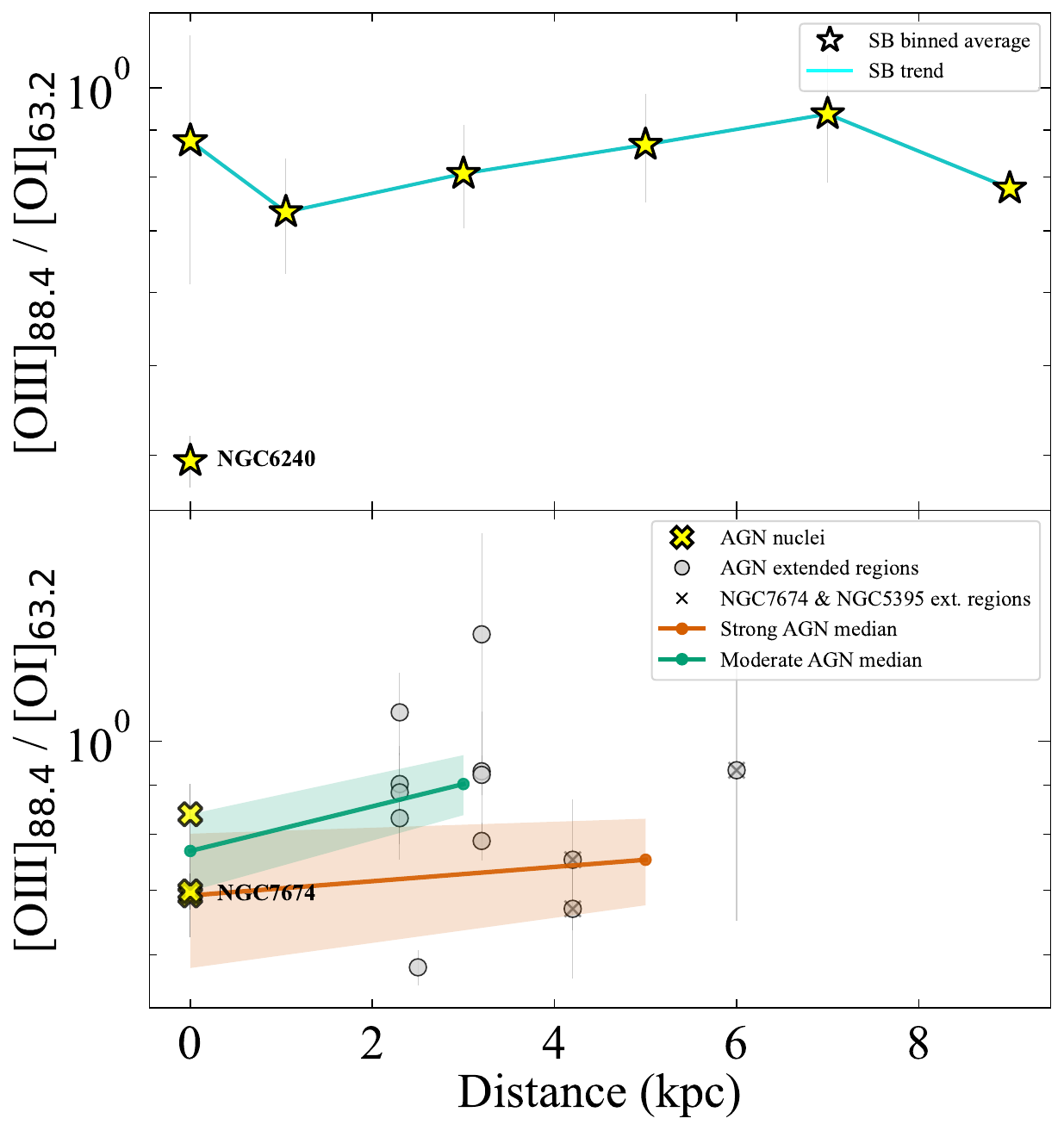}

    \caption{[Ar\textsc{iii}]$_{8.9}$/[Ar\textsc{ii}]$_{6.9}$
    (left panel) and [O\textsc{iii}]$_{88.3}$/[O\textsc{i}]$_{63}$
    (right panel) as a function of projected galactocentric distance.
    Symbols follow the same convention as in
    Fig.~\ref{fig:Ne_SIII_radial_trend}.}
    \label{fig:ArIII_OIII_radial_trends}
\end{figure}

\begin{figure}[H]
    \centering
    \includegraphics[width=0.90\textwidth]{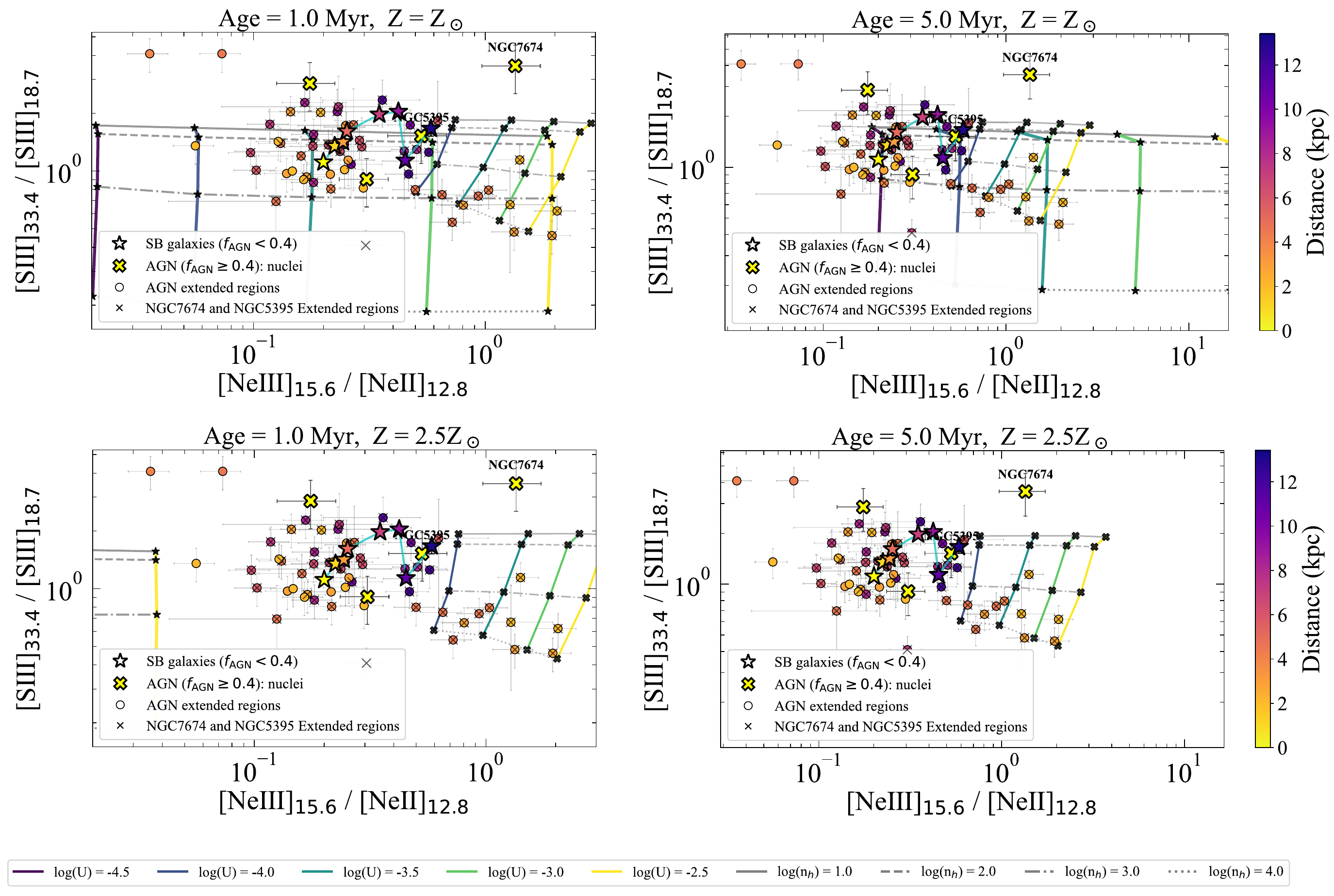}

    \caption{Additional CLOUDY photoionization model grids for the
    [S\textsc{iii}]$_{33}$/[S\textsc{iii}]$_{19}$ versus
    [Ne\textsc{iii}]$_{15}$/[Ne\textsc{ii}]$_{13}$ diagnostic diagram.
    The panels show models computed for stellar ages of 1 and 5 Myr
    and metallicities of $Z=Z_{\odot}$ and $Z=2.5Z_{\odot}$.
    Symbols follow the same convention as in Fig.~\ref{fig:SIII}.}
    \label{fig:SIII_appendix}
\end{figure}

\begin{figure}[H]
    \centering
    \includegraphics[width=0.90\textwidth]{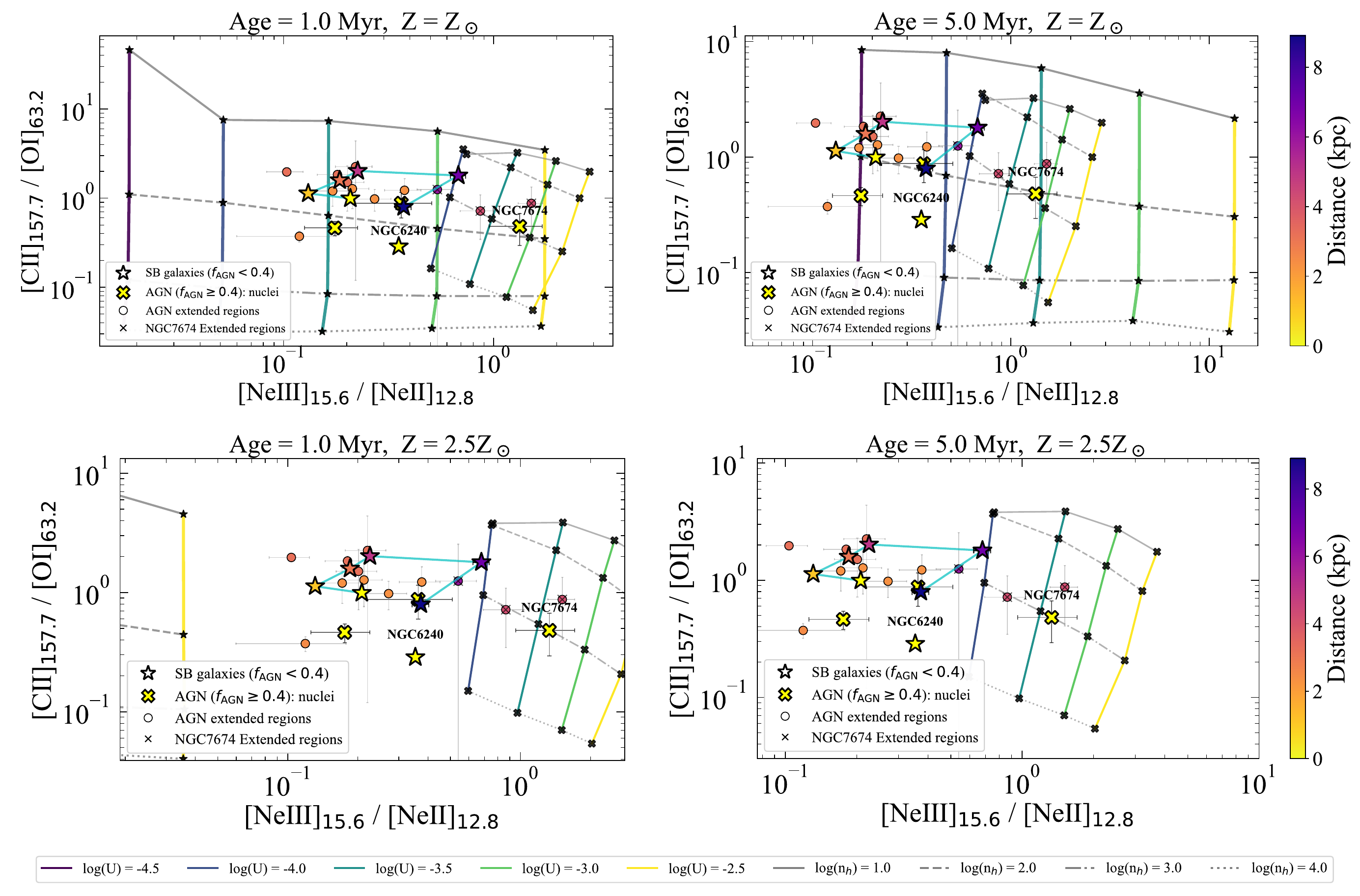}

    \caption{Additional CLOUDY photoionization model grids for the
    [C\textsc{ii}]$_{158}$/[O\textsc{i}]$_{63}$ versus
    [Ne\textsc{iii}]$_{15}$/[Ne\textsc{ii}]$_{13}$ diagnostic diagram.
    The panels show models computed for stellar ages of 1 and 5 Myr
    and metallicities of $Z=Z_{\odot}$ and $Z=2.5Z_{\odot}$.
    Symbols follow the same convention as in Fig.~\ref{fig:SIII}.}
    \label{fig:CII_appendix}
\end{figure}


\Needspace{0.30\textheight}
\section{\textit{Spitzer grid} radial trends}
\label{Spitzer models}

\begin{figure}[H]
    \centering

    \begin{minipage}[t]{0.32\textwidth}
        \centering
        \textbf{(a)}\\
        \includegraphics[width=\linewidth]
        {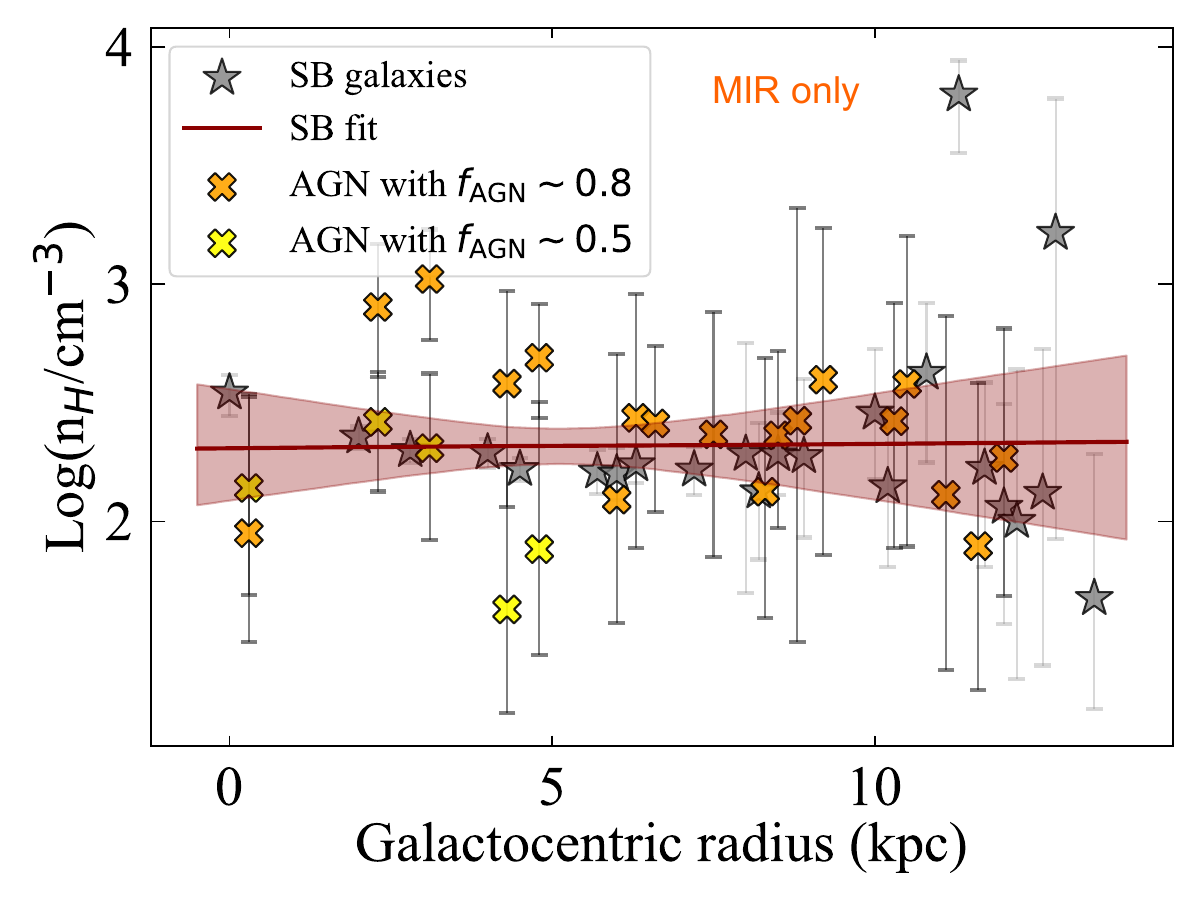}
    \end{minipage}
    \hfill
    \begin{minipage}[t]{0.32\textwidth}
        \centering
        \textbf{(b)}\\
        \includegraphics[width=\linewidth]
        {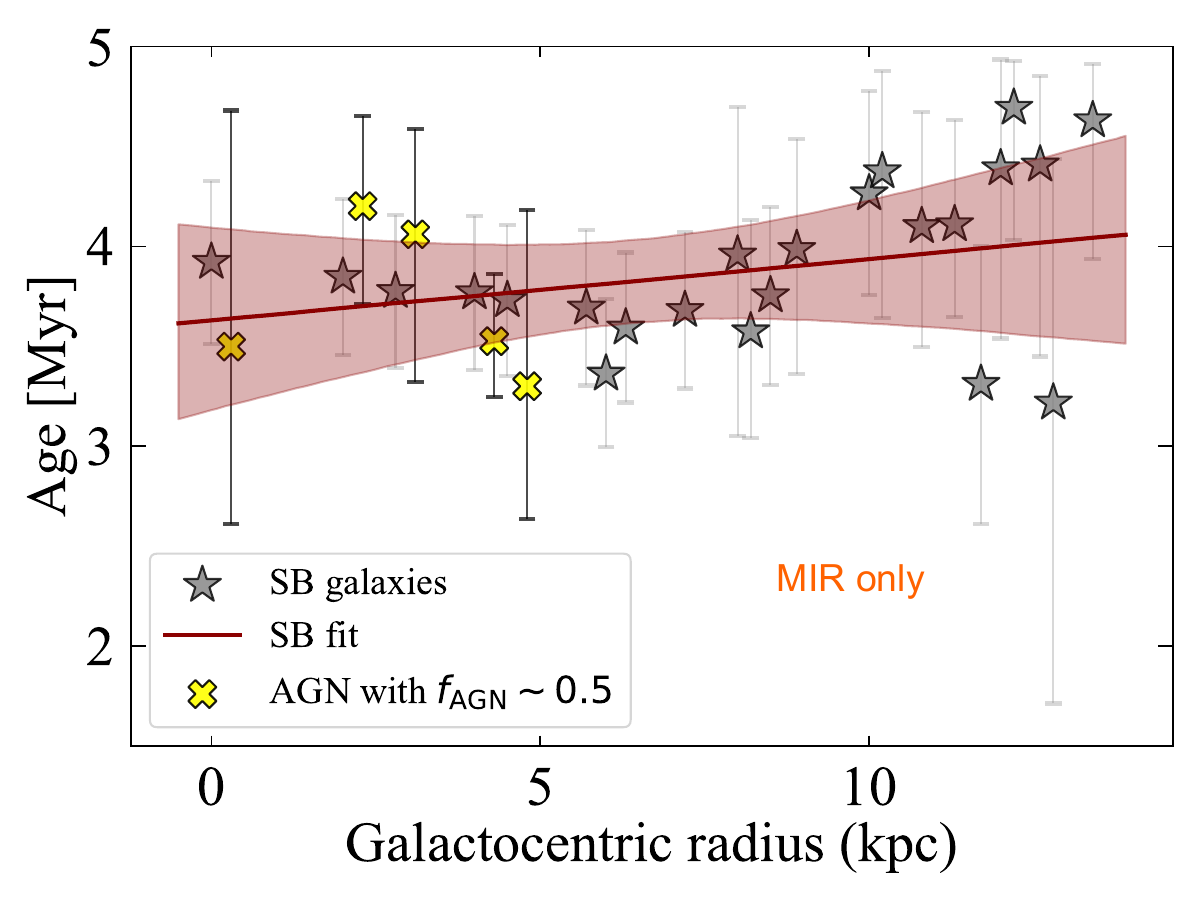}
    \end{minipage}
    \hfill
    \begin{minipage}[t]{0.32\textwidth}
        \centering
        \textbf{(c)}\\
        \includegraphics[width=\linewidth]
        {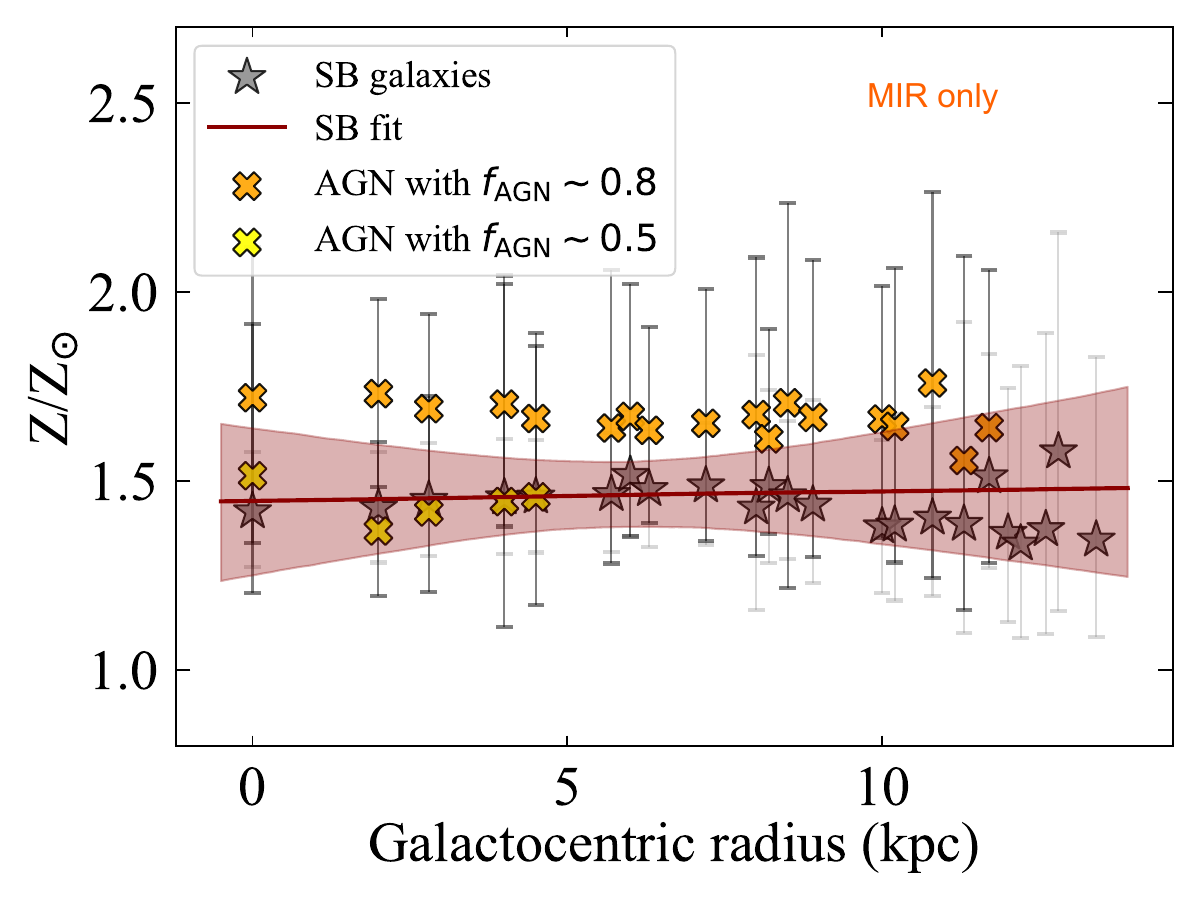}
    \end{minipage}

    \caption{Radial distributions of the physical parameters as
    functions of projected galactocentric distance for the
    \textit{Spitzer grid} configuration.
    Panel a: Radial profile of the hydrogen density,
    $\log(n_{\rm H})$.
    Panel b: Radial profile of the stellar age.
    Panel c: Radial profile of the gas metallicity.}
    \label{fig:physical_parameters_Spitzer}
\end{figure}

\end{appendix}

\end{document}